\documentclass[preprint,5p,times,twocolumn]{elsarticle}

\usepackage{amsmath, amsfonts, amssymb}
\usepackage{lineno}

\usepackage{newunicodechar,textgreek}  

\usepackage{algorithmic}
\usepackage{algorithm}
\usepackage{array}
\usepackage[acronym, toc]{glossaries}
\usepackage{siunitx}
\usepackage{makecell}

\usepackage{subcaption}

\usepackage{pifont}\newcommand{\xmark}{\ding{55}}

\usepackage[hidelinks]{hyperref}
\usepackage{nicematrix}
\usepackage{booktabs}

\journal{Signal Processing}

\makeglossaries
\glsenablehyper

\newacronym{allrad}{ALLRAD}{All-Round Ambisonic Decoding}
\newacronym{art}{ART}{Acoustic Ray Tracing}
\newacronym{cad}{CAD}{Computer Aided Design}
\newacronym{dirac}{DirAC}{Directional Audio Coding}
\newacronym{dmfa}{DMFA}{Dual Multiple Factor Analysis}
\newacronym{doa}{DOA}{Direction of Arrival}
\newacronym{dof}{DOF}{Degrees of Freedom}
\newacronym{ess}{ESS}{Exponential Sine Sweep}
\newacronym{fcr}{FCR}{Feedback Canceling Reverberator}
\newacronym{fdn}{FDN}{Feedback Delay Network}
\newacronym{fft}{FFT}{Fast Fourier Transform}
\newacronym{fir}{FIR}{Finite Impulse Response}
\newacronym{fwht}{FWHT}{Fast Walsh-Hadamard Transform}
\newacronym{gcc}{GCC}{Generalized Cross-Correlation}
\newacronym{gpu}{GPU}{Graphics Processing Unit}
\newacronym{hoa}{HOA}{Higher-Order Ambisonic}
\newacronym{hrtf}{HRTF}{Head-Related Transfer Function}
\newacronym{hws}{HWS}{Historical Worship Spaces}
\newacronym{iir}{IIR}{Infinite Impulse Response}
\newacronym{imu}{IMU}{Inertial Measurement Unit}
\newacronym{ir}{IR}{Impulse Response}
\newacronym{ism}{ISM}{Image-Source Method}
\newacronym{jnd}{JND}{Just Noticeable Difference}
\newacronym{lti}{LTI}{Linear Time-Invariant}
\newacronym{mdap}{MDAP}{Multiple-Direction Amplitude Panning}
\newacronym{mfa}{MFA}{Multiple Factor Analysis}
\newacronym{mls}{MLS}{Maximum Length Sequence}
\newacronym{nls}{NLS}{Nearest Loudspeaker Synthesis}
\newacronym{nupc}{NUPC}{Non-Uniform Partitioned Convolution}
\newacronym{ola}{OLA}{Overlap-Add}
\newacronym{ols}{OLS}{Overlap-Save}
\newacronym{pca}{PCA}{Principal Component Analysis}
\newacronym{rir}{RIR}{Room Impulse Response}
\newacronym{sdm}{SDM}{Spatial Decomposition Method}
\newacronym{sdn}{SDN}{Scattering Delay Network}
\newacronym{sfa}{SFA}{Sound Field Analysis}
\newacronym{shd}{SHD}{Spherical Harmonic Decomposition}
\newacronym{snr}{SNR}{Signal-to-Noise Ratio}
\newacronym{srir}{SRIR}{Spatial Room Impulse Response}
\newacronym{sti}{STI}{Speech Transmission Index}
\newacronym{vbap}{VBAP}{Vector Base Amplitude Panning}
\newacronym{vr}{VR}{Virtual Reality}
\newacronym{wfa}{WFA}{Wave Field Analysis}
\newacronym{wfs}{WFS}{Wave Field Synthesis}
\newacronym{ivp}{IVP}{Individual Vocabulary Profiling}
\newacronym{cvp}{CVP}{Consensus Vocabulary Profiling}
\newacronym{qda}{QDA}{Quantitative Descriptive Analysis}
\newacronym{fcp}{FCP}{Free Choice Profiling}
\newacronym{fp}{FP}{Flash Profile}
\newacronym{gpa}{GPA}{Generalized Procrustes Analysis}
\newacronym{foa}{FOA}{First-Order Ambisonics}
\newacronym{svd}{SVD}{Singular Value Decomposition}
\newacronym{rt}{RT}{Reverberation Time}
\newacronym{edt}{EDT}{Early Decay Time}
\newacronym{icc}{ICC}{Interaural Cross-Correlation}
\newacronym{asw}{ASW}{Apparent Source Width}
\newacronym{drr}{DRR}{Direct-to-Reverberant Ratio}
\newacronym{ail}{AIL}{Alamire Interactive Lab}
\newacronym{sfs}{SFS}{Sound Field Synthesis}
\newacronym{mushra}{MUSHRA}{MUltiple Stimuli with Hidden Reference and Anchor}
\newacronym{aa}{AA}{Acoustic Acquisition}
\newacronym{rta}{RTA}{Real-Time Auralization}
\newacronym{pe}{PE}{Perceptual Evaluation}
\newacronym{fvn}{FVN}{Filtered Velvet Noise}
\newacronym{gan}{GAN}{Generative Adversarial Network}
\newacronym{naf}{NAF}{Neural Acoustic Field} 

\usepackage{titlesec}

\titleformat{\paragraph}
{\normalfont\normalsize\itshape}{\theparagraph}{1em}{}\titlespacing*{\paragraph}{0pt}{3.25ex plus 1ex minus .2ex}{1.5ex plus .2ex}

\begin{document}
\begin{frontmatter}

\title{A State-of-the-Art Review on Acoustic Preservation of Historical Worship Spaces through Auralization}

\author[]{Hannes Rosseel\corref{cor1}} \cortext[cor1]{Corresponding author.}
\ead{hannes.rosseel@esat.kuleuven.be}
\author[]{Toon van Waterschoot\corref{cor2}}
\cortext[cor2]{EURASIP Member}

\affiliation{
    organization={KU Leuven, Department of Electrical Engineering (ESAT), STADIUS Center for Dynamical Systems, Signal Processing and Data Analytics},
    addressline={Kasteelpark Arenberg 10},
    city={Leuven},
    postcode={B-3001},
    state={Vlaams-Brabant},
    country={Belgium}}

  \begin{abstract}
   \gls{hws} are significant architectural landmarks which hold both cultural and spiritual value. The acoustic properties of these spaces play a crucial role in historical and contemporary religious liturgies, rituals, and ceremonies, as well as in the performance of sacred music. However, the original acoustic characteristics of these spaces are often at risk due to repurposing, renovations, natural disasters, or deterioration over time. This paper presents a comprehensive review of the current state of research on the acquisition, analysis, and synthesis of acoustics, with a focus on \gls{hws}. An example case study of the Nassau chapel in Brussels, Belgium, is presented to demonstrate the application of these techniques for the preservation and auralization of historical worship space acoustics. The paper concludes with a discussion of the challenges and opportunities in the field, and outlines future research directions.
\end{abstract}
\glsreset{hws} 

\begin{keyword}
Acoustic preservation \sep historical worship spaces \sep room acoustics \sep auralization
  \end{keyword}
\end{frontmatter}

\section{Introduction}
\label{sec:review2024:introduction}
\noindent

\gls{hws} form a significant part of the architectural heritage of many cultures, and are often considered to be architectural landmarks. These spaces provide a unique acoustic environment, mainly characterized by long reverberation times, that is integral to the performance of religious rituals, ceremonies, and the performance of sacred music \cite{giron2017church}. The original acoustic properties of \gls{hws} are often at risk due to repurposing, renovations, natural disasters, or deterioration over time \cite{katz2020exploring, katz2020past}. It is therefore important to the cultural and spiritual communities, as well as to the wider public, to safeguard and preserve the aural heritage that these spaces provide. The safeguarding of intangible cultural heritage has been recognized by UNESCO \cite{unesco2022basic}.

Existing review papers on the acoustics of \gls{hws} have focused mainly on the historical development of these spaces, the acoustic properties of \gls{hws}, and the impact of architectural design decisions on the acoustics of these spaces \cite{giron2017church, zhang2024soundscape}. However, there is a lack of comprehensive reviews on the state of research on the preservation, analysis, and reproduction of \gls{hws} acoustics. To address this literature gap, this paper presents a comprehensive review of the current state of research on acoustic preservation and acoustic reproduction of \gls{hws} through auralization. Auralization is the process of rendering an acoustic environment audible through the use of sound reproduction techniques \cite{vorlander2020auralization}. Auralization has been widely used in the field of architectural acoustics to simulate the acoustic properties of spaces, and to evaluate the impact of design decisions on the acoustic performance of buildings \cite{vorlander2020auralization}. In the context of \gls{hws}, auralization can be used to safeguard and recreate the original acoustic characteristics of a space, and to assess the impact of alterations or renovations on the acoustic environment \cite{katz2020exploring}.

\begin{figure*}
    \centering
    \glsreset{hws}
    \includegraphics[width=\linewidth]{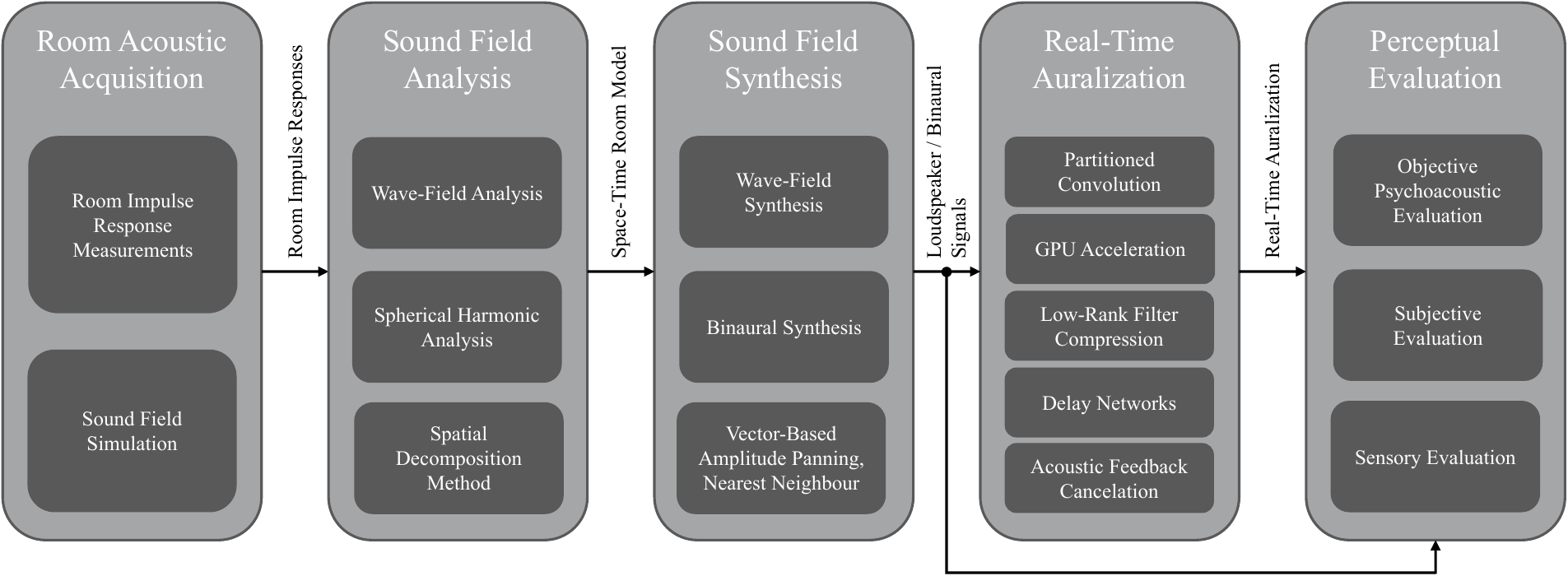}
    \caption{Overview of an auralization process for the preservation and reproduction of \gls{hws}. The process consists of five main components: room acoustic acquisition, sound field analysis, sound field synthesis, real-time auralization, and perceptual evaluation. The dark-colored boxes represent the relevant methods for each component. The components and their corresponding methods can be utilized either in combination or individually as tools to preserve and reproduce the acoustics of \gls{hws}. All components and methods, together with their application to the preservation of historical worship space acoustics, are discussed in detail in the following sections.}
    \label{fig:review2024:overview}
\end{figure*}

An overview of the key components in the auralization process for \gls{hws} is provided in Fig. \ref{fig:review2024:overview}. This process can be divided into five main components: room \gls{aa}, \gls{sfa}, \gls{sfs}, \gls{rta}, and \gls{pe}. The dark-colored boxes represent the relevant methods for each component. The components and their corresponding methods can be utilized either in combination or individually as tools to preserve and reproduce the acoustics of \gls{hws}.

The structure of this paper is organized as follows. In Section \ref{sec:review2024:summary_hws}, a literature review on the acoustic preservation of \gls{hws} through auralization is presented, situating the relevant research within the framework of the five main components of the auralization process. Following this, each individual component is discussed in detail from Sections \ref{sec:review2024:room_acoustic_acquisition} to \ref{sec:review2024:perceptual_evaluation}, with a focus on the current state of research and its application to the preservation and reproduction of \gls{hws}. Section \ref{sec:review2024:case_study} presents a case study on the preservation and auralization of the Nassau Chapel in Brussels, Belgium. Finally, the paper concludes in Section \ref{sec:review2024:conclusion} with a discussion of the challenges and opportunities in the field, and an outline of future research directions.

 \section{Acoustic Preservation of Historical Worship Spaces through Auralization}
\label{sec:review2024:summary_hws}
In this section, an overview of literature relating to the preservation through auralization of \gls{hws} is provided. The overview provides a summary of research conducted in \gls{hws}, focusing on key components of the auralization process: room acoustic acquisition, sound field analysis, sound field synthesis, real-time auralization, and perceptual evaluation. 

The results of the literature search are summarized in Table \ref{tab:review2024:overview}. This table provides an overview of research conducted in \gls{hws} where the focus was on the analysis, preservation and reproduction of the acoustics of the spaces through auralization. The table includes the author(s) and publication date, type of worship space, and the employed methods: \gls{aa}, \gls{sfa}, \gls{sfs}, \gls{rta}, and \gls{pe}. The studies are ordered by publication date. A brief summary of each study is provided below.\\[3pt]

\glsunsetall
\begin{table*}
  \centering
  \resizebox{\textwidth}{!}{\rotatebox{90}{\begin{minipage}{\textheight}
        \glsreset{aa}
        \glsreset{sfa}
        \glsreset{sfs}
        \glsreset{rta}
        \glsreset{pe}      
        \caption{Summary of research relating to preservation through auralization of historical worship space acoustics. The table includes the author(s) and publication date, type of worship space, \gls{aa} method, \gls{sfa} method, \gls{sfs} method, \gls{rta} method, and \gls{pe} method. The acronyms used in this table are defined in the glossary at the end of the paper. Studies are ordered by publication date.}
        \label{tab:review2024:overview}
        \begin{NiceTabular}{ X[l,7] X[l,4] X[l,4] X[l,2] X[l,5] X[l,5] X[l,5] }
          \toprule
          Author(s) & Worship Space & AA & SFA & SFS & RTA & PE \\
          \midrule
Weitze \textit{et al.} (2003) \cite{weitze2003comparison} & Mosque, Byzantine church & ODEON & \xmark & Binaural & \xmark & \xmark \\
Koutsivitis \textit{et al.} (2005) \cite{koutsivitis2005reproduction} & Greek temple & \gls{art} & \xmark & Binaural & partitioned \gls{ols} & \xmark \\
Martellotta (2008) \cite{martellotta2008subjective} & Catholic church & \gls{ess} & \xmark & Stereo loudspeaker & \xmark & Preference-based listening test \\
Fazenda and Drumm (2013) \cite{fazenda2013recreating} & Stonehenge & \makecell[l]{Balloon bursts, \\\gls{ess}} & \xmark & \gls{wfs}, \gls{foa} & \xmark & \xmark \\
Pedrero \textit{et al.} (2014) \cite{pedrero2014virtual} & Pre-Romanesque churches & RAVEN & \xmark & Binaural & \xmark & \xmark \\
Postma and Katz (2015) \cite{postma2015creation} & Abbey church & CATT-Acoustic, \gls{ess} & \xmark & \xmark & \xmark & Objective parameters \\
Elicio and Martellotta (2015) \cite{elicio2015acoustics} & Orthodox church & \gls{ess} & \gls{shd} & \xmark & \xmark & Objective parameters \\
Frank and Zotter (2016) \cite{frank2016spatial} & Church & \xmark & \gls{sdm}, \gls{shd} & Ambisonics, Loudspeaker array & \xmark & Listening tests \\
Postma \textit{et al.} (2016) \cite{postma2016virtual} & Abbey church, cathedral & CATT-Acoustic & \gls{shd} & Ambisonics, Binaural & Head Tracking, Max/MSP & Objective parameters \\
Postma and Katz (2016) \cite{postma2016perceptive} & Abbey church & \gls{ess}, CATT-Acoustic & \xmark & Binaural & \xmark & Comparative listening tests \\
Alvarez-Morales \textit{et al.} (2017) \cite{alvarezmorales2017virtual} & Cathedral & \gls{ess}, CATT-Acoustic & \xmark & Binaural & \xmark & Perceptual attribute ranking \\
Martellotta \textit{et al.} (2018) \cite{martellotta2018investigation} & Cathedral and Crypt & \gls{ess} & \gls{shd} & \xmark & \xmark & Objective parameters \\
Su{\'a}rez \textit{et al.} (2018) \cite{suarez2018virtual} & Hypostyle mosque & \gls{ess}, CATT-Acoustic & \xmark & Binaural & \xmark & Objective parameters \\
Canfield-Dafilou \textit{et al.} (2019) \cite{canfielddafilou2019method} & Catholic church & All-pass chirp & \xmark & Loudspeaker array & \gls{fcr} & Interactive evaluation \\
Katz \textit{et al.} (2020) \cite{katz2020acoustic} & Cathedral & \gls{ess}, Starter pistol, Balloon bursts & \gls{sdm}  & \xmark & \xmark & \xmark \\
{\'A}lvarez-Morales \textit{et al.} (2020) \cite{alvarezmorales2020acoustic} & Cathedral & \gls{ess}, ODEON & \gls{shd} & \xmark & \xmark & Objective parameters \\
Autio \textit{et al.} (2021) \cite{autio2021historically} & Abbey & ODEON & \xmark & Binaural & \xmark & Objective parameters \\
Eley \textit{et al.} (2021) \cite{eley2021virtual} & Cathedral & CATT-Acoustic & \gls{shd} & \makecell[l]{Ambisonics, Binaural,\\Loudspeaker array} & \gls{fdn}, \gls{nupc}, Max/MSP & Interactive evaluation \\
Warusfel and Emerit (2021) \cite{warusfel2021assessing} & Egyptian temple & \gls{ess} & \gls{shd} & Ambisonics, Binaural & Max/MSP & \xmark \\
Berger \textit{et al.} (2023) \cite{berger2023exploring} & Catholic church and mosque & \xmark & \gls{shd} & Ambisonics, Binaural & \gls{vr} 6-\gls{dof} Head-tracking & \xmark \\
Mullins and Katz (2023) \cite{mullins2023immersive} & Cathedral & CATT-Acoustic & \xmark & Binaural & Partitioned convolution, Max/MSP & Interactive, Preference \\
\bottomrule
        \end{NiceTabular}
      \end{minipage}
    }
  }
  \end{table*}

\glsresetall
\glsunset{hws}

\noindent The Conservation of the Acoustical Heritage and Revival of Sinan's mosques Acoustics (CAHRISMA) project was a collaborative research project that aimed to study the acoustics of historical mosques and Byzantine churches built by the architect Sinan in Istanbul, Turkey. The project aimed to study the acoustics of the spaces using a combination of measurement and simulation techniques. Acoustic measurements were conducted in three mosques and four Byzantine churches by capturing the \glspl{rir} using both a binaural and B-format probe at various locations in the spaces using the \gls{ess} method. In a different study from the same project, the previously mentioned in-situ recordings of three spaces were compared to auralizations obtained from geometrical models of the spaces using the ODEON software \cite{weitze2003comparison}. The comparison was made by convolving the recorded binaural \glspl{rir} of a space with anechoic recordings and comparing the results to the auralizations obtained from the ODEON model. The study found that the auralizations were able to reproduce the acoustics of the spaces with a good resemblance to the measured data \cite{weitze2003comparison}.

Koutsivitis \textit{et al.} \cite{koutsivitis2005reproduction} presented a methodology for real-time interactive auralization and visualization techniques for reproducing real world, real-time events in open and closed spaces in 2005. The methodology was applied to the real-time reproduction of the acoustics of the ancient theater of Epidauros, and the ancient temple of Zeus, which are both located in Greece. The acoustics of the spaces were simulated from a geometrical model using \gls{art} to obtain sets of binaural \glspl{rir} for each space. By using readily available \glspl{hrtf} and combining them with the simulated binaural \glspl{rir}, the acoustics of the spaces were auralized by tracking the position of the listener in the virtual space and updating the binaural \glspl{rir} accordingly. The auralization was performed in real-time by calculating the convolutions of the binaural \glspl{rir} with the input signal using a partitioned \gls{ols} algorithm \cite{koutsivitis2005reproduction}.

A subjective study on preferred listening conditions in Italian Catholic churches was performed by Martellotta in 2008 \cite{martellotta2008subjective}. The acoustics of nine Italian Catholic churches were measured using the \gls{ess} method as excitation signal and a B-format microphone and a binaural head and torso. The measured \glspl{rir} were convolved with anechoic musical excerpts and played back to subjects over loudspeakers in a 'stereo dipole' configuration in a purpose-built dry room. A total of 143 subjects took part in the listening tests, in which they were presented with paired comparisons of the different recordings and asked to state their preference. The results of the listening tests were then analyzed using factor analysis to determine the key acoustic parameters that influenced the subjects' preferences and to determine optimal acoustic conditions for different types of music \cite{martellotta2008subjective}.

Fazenda and Drumm \cite{fazenda2013recreating} presented a study on the recreation of the sound of Stonehenge using \gls{wfs} in 2013. The study first measured the acoustics of the original Stonehenge monument in Wiltshire, England using balloon bursts as the excitation signal. These measurements were compared to measurements made with an omnidirectional and a first-order Ambisonic microphone at a restored replica of Stonehenge located in Maryhill, Washington, USA. The acoustics measured at the replica were then used to recreate the acoustics of the original Stonehenge monument using a combination of \gls{wfs} and First-Order Ambisonic synthesis \cite{fazenda2013recreating}.

Pedrero \textit{et al.} presented a study in 2014 on the virtual recovery of the sound of the Hispanic rite \cite{pedrero2014virtual}. The study created geometric models of five pre-Romanesque churches in the Iberian Peninsula using historical sources to reconstruct the buildings in their primitive state. The acoustic simulation software RAVEN \cite{schroder2011raven} was used to simulate the acoustics of the spaces for static and moving source and receiver positions. A total of eight musical pieces were recorded under anechoic conditions and auralized in the acoustic models of the churches \cite{pedrero2014virtual}.

Postma and Katz \cite{postma2015creation} developed a method for calibrating geometrical acoustic models with acoustical measurements. The method was applied to a geometric model of the abbey church Saint-Germain-des-Pr{\'e}s in Paris, France. The acoustics of the space was simulated using the CATT-Acoustic software, and the simulated acoustics was calibrated by altering the absorption and scattering coefficients of the surfaces in the model to match the acoustical measurements made in the spaces. The calibrated acoustic model was then used as a basis for simulating the acoustics of the space as it would have sounded in the 17th century. Historical sources were used to recreate a configuration of the space that was historically accurate in the context of liturgical practices of the time. The study found that the calibrated acoustic model provided a plausible representation of the acoustics of the space as it would have sounded in the 17th century \cite{postma2015creation}.

The acoustics of the Orthodox church of San Nicola in Bari, Italy, was studied by Elicio and Martellotta \cite{elicio2015acoustics}. The acoustics were measured at different positions in the church using the \gls{ess} method with a dodecahedron loudspeaker and an \textit{Eigenmike 32} spherical microphone array. The measurements were analyzed in terms of key acoustic parameters. A spatial analysis of the sound field was performed using higher-order \gls{shd} to identify the spatial distribution of the sound energy in the church, and creating a directional reflection map of the church \cite{elicio2015acoustics}.

Frank and Zotter presented a study in 2016 that examines spatial impression and directional resolution of synthesized sound fields reproduced with an Ambisonic loudspeaker playback system \cite{frank2016spatial}. The \glspl{rir} utilized in this study were measured in St. Andrew's church, Lyddington, UK, and sourced from the openAIR database \cite{openair2024}. The study employed the \gls{sdm} to enhance the first-order B-format \glspl{rir} to higher spherical harmonic orders. Acoustic reproduction was achieved through Ambisonic decoding of the spherical harmonics to a loudspeaker array. The reproduction was conducted for the original first-order, the \gls{sdm}-enhanced first-order, the \gls{sdm}-enhanced third-order, and the \gls{sdm}-enhanced fifth-order representation of the sound field. A perceptual evaluation of the auralizations using the different representations was performed with two listening tests. The results suggest that higher-order Ambisonics playback provides a larger sweet spot with a more convincing reverberant sound field in terms of envelopment and depth \cite{frank2016spatial}.

A study by Postma \textit{et al.} \cite{postma2016virtual} presented a performance by the Conservatoire de Paris using virtually reconstructed acoustics of the Cathédrale Notre-Dame de Paris. The virtual acoustics were reconstructed based on a geometrical model of the cathedral using the CATT-Acoustic software, and calibrated using acoustic measurements made in the cathedral. The performance was auralized using binaural Ambisonic synthesis and head tracking to provide an immersive experience of the acoustics of the cathedral. The study found that the virtual acoustics provided a realistic representation of the acoustics of the cathedral \cite{postma2016virtual}.

The achievable perceptual quality of room acoustic simulations and auralizations has been studied by Postma and Katz \cite{postma2016perceptive}. Binaural auralizations were created from simulations of the acoustics of an abbey church Saint-Germain-des-Pr{\'e}s, the Cath{\'e}drale Notre-Dame de Paris, and the Th{\'e}{\^a}tre de l'Ath{\'e}n{\'e}e in Paris, France. The acoustic simulations were calibrated from \gls{ess} measurements of the spaces. The study carried out subjective listening tests to compare the perceptual similarity of simulated and measured auralizations. The participants were asked to rate the similarity of the auralizations using eight predefined perceptual attributes. 

Álvarez-Morales \textit{et al.} explored the use of \gls{vr} to create an immersive acoustic reconstruction of the Royal Chapel of the Gothic Cathedral of Seville, Spain \cite{alvarezmorales2017virtual}. A simplified geometric model of the Cathedral of Seville and the Royal Chapel was created in \gls{cad} software, and used to simulate the acoustics of the space using CATT-Acoustic. The acoustic simulation was calibrated using acoustical measurements made in the Royal Chapel using the \gls{ess} method. A \gls{vr} auralization system was developed that combined the acoustic auralization with a static visual representation of the space. The acoustic auralizations were performed using binaural synthesis, and the visual representation of the space was displayed using a projector. A preliminary listening test was carried out to perceptually evaluate the experience of the \gls{vr} auralization system. The perceptual evaluation was conducted using a questionnaire that evaluated the realism of the acoustic auralization and the visual representation of the space. The participants had to rank the acoustic experience for different stimuli \cite{alvarezmorales2017virtual}.

In Martellotta \textit{et al.} \cite{martellotta2018investigation}, an acoustic analysis of the Cathedral of Cadiz and the underlying crypt, which is acoustically connected to the main cathedral, is presented. The crypt, with its central vaulted rotunda connected to radial galleries and openings to the main cathedral, exhibits a combination of acoustic phenomena like flutter echoes and non-diffuse sound propagation \cite{martellotta2018investigation}. The acoustics of the cathedral and the crypt were measured using a first-order spherical microphone with the \gls{ess} method. The measurements were analyzed by calculating the reverberation time and \gls{edt} for different octave bands, and Bayesian analysis was performed to determine the decay rate of the \glspl{rir}. The sound field distribution was visualized using directional intensity maps. The study found that the crypt's geometry leads to a non-diffuse sound field in the rotunda, with reflections between the floor and dome causing flutter echoes and longer reverberation times \cite{martellotta2018investigation}.

Su{\'a}rez \textit{et al.} presented a study on the virtual acoustic reconstruction of the hypostyle mosque of Córdoba, located in Spain \cite{suarez2018virtual}. An acoustic simulation was made from geometric models of the mosque using CATT-Acoustic software. The acoustic simulation was calibrated using acoustical measurements made in the mosque using the \gls{ess} method. The calibrated acoustic model was used to simulate binaural auralizations of the mosque \cite{suarez2018virtual}.

A method for studying the relationship between musical composition and room acoustics of the Chiesa di Sant'Aniceto in Rome, Italy, was presented by Canfield-Dafilou \textit{et al.} \cite{canfielddafilou2019method}. The study used a six-second all-pass chirp signal to measure a set of \glspl{rir} of the Catholic church at positions where the singers were likely to have been located. The processing of the \glspl{rir} consisted of removing the direct sound, and generating sets of \glspl{rir} with different reverberation times. An interactive auralization system was developed using a loudspeaker array and the \gls{fcr} method \cite{abel2018feedback} to reproduce the acoustics of the church in real-time with different reverberation times. Participants evaluated the auralizations by first interacting with the auralization system by performing a piece of vocal music for all reverberation times, and then by evaluating the auralizations using questionnaires, which evaluated the preference of the participants for the different reverberation times \cite{canfielddafilou2019method}.

Following the destructive fire that occurred in Cathédrale Notre-Dame de Paris in 2019, the acoustic properties of the cathedral were found to have been altered by comparing pre-fire acoustic measurements from 1987 and 2015 to post-fire measurements made on the construction site in 2020 \cite{katz2020acoustic}. The 1987 measurements were recorded using balloon bursts as the excitation signal, while the 2015 measurements were recorded at roughly the same location using sine sweeps as the excitation signal \cite{katz2020acoustic}. The novel post-fire measurements were made using a 20-second-long sine sweep, which was played back through a loudspeaker and recorded at the several receiver positions using a microphone, as detailed in \cite{katz2020acoustic}. These measurements were complemented by measurements made using starter pistol gun shots and balloon bursts. After the measurements were made, a spatial analysis of the sound field was performed using \gls{sdm} to identify the spatial distribution of the sound energy in the cathedral \cite{katz2020acoustic}. The study found that there was a significant decrease in the reverberation time of the cathedral after the fire.

Álvarez-Morales \textit{et al.} presented an acoustic survey of the York Minster's Chapter House, located in York, UK. The acoustics of the Chapter House were measured at several positions using the \gls{ess} method and captured using a Soundfield ST450 microphone, and a binaural dummy head. A geometric model of the space was created using \gls{cad} software, and the acoustics of the space were simulated using ODEON software. The simulated acoustics were calibrated using the acoustical measurements made in the Chapter House. A perceptual analysis of the simulations was performed by calculating objective acoustic parameters \cite{alvarezmorales2020acoustic}.

Autio \textit{et al.} \cite{autio2021historically} performed a historically accurate virtual acoustic reconstruction of the Vadstena Abbey church during a 15th-century service. The church's original acoustics were lost due to renovations. A three-dimensional model of the church was created based on historical sources, and \gls{art} was used to simulate the acoustic model in ODEON software. To improve the quality of the simulated acoustics, the absorption coefficients of several surfaces were calibrated using acoustic measurements made in the church. The virtual acoustic auralization was performed using binaural synthesis \cite{autio2021historically}. The study found that the virtual acoustic reconstruction provided a historically accurate representation of the acoustics of the church during a 15th-century service \cite{autio2021historically}.

A study by Eley \textit{et al.} presented an interactive real-time auralization system for reproducing Cath{\'e}drale Notre-Dame de Paris' acoustics using real-time binaural and multichannel loudspeaker synthesis \cite{eley2021virtual}. The aim of the study was to perceptually validate the virtual reconstruction of the cathedral's acoustics prior to the 2019 fire. Real-time auralization of acoustics was achieved by reproducing the cathedral's acoustic \glspl{rir} in real-time using a hybrid approach which combined non-uniform partitioned convolution for reproducing the early reflections, and \gls{fdn} for reproducing the late reverberation. Four singers, members of a medieval ensemble with experience performing in the Cath{\'e}drale Notre-Dame de Paris, were asked to validate the reproduction system by performing excerpts of medieval repertoire in the virtual reproduction space in both binaural and multichannel loudspeaker configurations. The singers' feedback was collected using questionnaires which evaluated the similarity of the virtual acoustics to the real acoustics of the cathedral \cite{eley2021virtual}. It should be noted that the participants could only base their comparison on their memory of the cathedral's acoustics \cite{eley2021virtual}, which affects the reliability of the results.

Warusfel and Emerit \cite{warusfel2021assessing} presented a study on the acoustic assessment of the Dendara temple, located in Egypt. The acoustics of the temple were measured at several positions using the \gls{ess} method, and captured using an \textit{Eigenmike 32} spherical microphone array. A spatial analysis was performed using fourth-order \gls{shd} to identify the energy decay deviation of the sound energy in the temple. The measurements were used to auralize the acoustics of the edifice using real-time binaural \gls{hoa} synthesis using the Max/MSP software. In future research, the measurements will be used to calibrate a geometrical acoustic model of the temple to simulate the acoustics of the space as it would have sounded in antiquity \cite{warusfel2021assessing}.

In \cite{berger2023exploring}, Berger \textit{et al.} created a \gls{vr} auralization system for exploring the acoustics of the Jaame-Abbasi mosque (Shah mosque) in Isfahan, Iran, and the Santa Maria dei Derelitti church in Venice, Italy. The acoustics of the spaces were measured using a first-order Ambisonic microphone. These measurements, together with a three-dimensional model of the spaces, were used to create a \gls{vr} auralization system that allowed users to explore the acoustics of the spaces in real-time using a six-\gls{dof} head-tracking \gls{vr} system. The study found that the \gls{vr} auralization system provided an immersive experience of the acoustics of the spaces, and allowed users to explore the acoustics of the spaces from different perspectives \cite{berger2023exploring}.

To better understand the relationship between ancient music performance practices and the acoustics of the performance space, an immersive binaural auralization system was developed and presented by Mullins and Katz in 2023 \cite{mullins2023immersive}. The system was designed to provide a binaural auralization of the Cath{\'e}drale Notre-Dame de Paris' acoustics during choral performances. A set of binaural \glspl{rir}, which model the acoustics of the cathedral, were simulated using CATT-Acoustic. The binaural synthesis used a set of \glspl{hrtf} to achieve realistic spatialization of the reproduced sound. The auralization system was designed to be used in real-time by efficiently convolving the binaural synthesis filters with the input signal using partitioned convolution methods. A calibration process ensured that the latency and energy ratios of the auralization system were within acceptable limits. A subjective analysis in the form of an interactive choir performance was conducted to evaluate the perceptual quality of the auralization \cite{mullins2023immersive}.\\[3pt]

In the following sections, we provide a detailed overview of the research conducted in \gls{hws}, focusing on key components of the auralization process: room acoustic acquisition, sound field analysis, sound field synthesis, real-time auralization and perceptual evaluation. Each section provides a general overview of the research methods, after which a detailed focus is given to research that is relevant or has been applied to the acoustic preservation and reproduction of \gls{hws}.

\glsresetall
\glsunset{hws} \section{Room Acoustic Acquisition}
\label{sec:review2024:room_acoustic_acquisition}

\subsection{Methods}
\label{sec:review2024:room_acoustic_acquisition_methods}

The acoustic properties of a room are determined by its geometry, surface materials, and the presence of sound-absorbing objects. From a system identification perspective, the acoustics between a sound source and a receiver positioned within the enclosure can be modeled as a \gls{lti} system, assuming geometric and atmospheric conditions in the room remain constant over time \cite{vorlander1997practical, prawda2023time}. The behavior of an \gls{lti} system is characterized entirely by its \gls{ir}. For acoustics in a room, this is referred to as the \gls{rir}, which represents the acoustic response of the room to an impulsive sound source, captured at a receiver position. A \gls{rir} $h(t, \mathbf{x}, \mathbf{r})$ is a function of time $t$, the source position $\mathbf{x}$, and the receiver position $\mathbf{r}$, and can be modeled as the sum of all acoustic events in the room \cite{tervo2013sdm}:

\begin{equation}
    \label{eq:review2024:rir_events}
    h(t, \mathbf{x}, \mathbf{r}) = \sum_{p=0}^{\infty} h_p(t, \mathbf{x}_p, \mathbf{r}),
\end{equation}

\noindent where $h_p(t, \mathbf{x}_p, \mathbf{r})$ is the impulse response of the $p$-th event, where $\mathbf{x}_p$ denotes the $p$-th event's position.

In order to measure the acoustics of a given space, it is necessary to obtain the \glspl{rir} between relevant source and a receiver positions within the space. In spaces that are accessible to researchers, the \gls{rir} between a source and receiver position can be acquired by measuring the response of the room to an impulsive sound source, following ISO 3382 \cite{ISO3382-1, ISO3382-2}. For inaccessible or acoustically altered spaces, the \gls{rir} can be estimated from the room's geometry and surface material properties using simulation methods, such as the \gls{ism}, \gls{art}, numerical wave-based methods, statistical methods, or using a combination of those. This section provides an overview of the methods used to acquire \glspl{rir} in accessible and inaccessible spaces, and discuss the challenges associated with acquiring \glspl{rir} in \gls{hws}.

\subsubsection{Room Impulse Response Measurements in Accessible Spaces}

The \gls{rir} in spaces that are still accessible can be acquired by exciting the room with an omnidirectional impulsive sound source, and capturing the response at a receiver position \cite{kuttruff2017room}. Prior to the advent of digital signal processing, the most common methods for acquiring \glspl{rir} were typically conducted through the use of hand claps, balloon pops \cite{griesinger1996beyond, horvat2008comparison, patynen2011investigations}, high-voltage sparks \cite{lamothe1985acoustical}, or by firing a starter pistol \cite{lamothe1985acoustical, griesinger1996beyond} in the room, and recording the acoustic response using a tape recorder. However, these methods lack repeatability and exhibit poor omnidirectionality, which can lead to inaccuracies in the measured \glspl{rir} \cite{griesinger1996beyond}.

To achieve accurate and repeatable \gls{rir} measurements, a space can be excited with a known signal $s(t)$, that is played back through a loudspeaker system. The acoustic response $y(t, \mathbf{x}, \mathbf{r})$ can then be captured at a receiver position $\mathbf{r}$ using a microphone setup \cite{kuttruff2017room}. Crucially, the excitation signal $s(t)$ should exhibit an autocorrelation function that is a Dirac delta function, i.e., $s(t) \ast s(-t) = \delta(t)$, where $\delta(t)$ is the Dirac delta function and $(\ast)$ denotes the convolution operator. The \gls{rir} can then be estimated by convolving the response signal with the time-reversed excitation signal, as

\begin{equation}
    h(t, \mathbf{x}, \mathbf{r}) = y(t, \mathbf{x}, \mathbf{r}) \ast s(-t).
\end{equation}

A common excitation signal that meets the above criteria is the \gls{mls} signal, proposed for measuring room acoustics by Schroeder \cite{schroeder1979integrated}, and further researched in \cite{borish1983efficient, rife1989transfer, dunn1993distortion, vanderkooy1994aspects, novak2016impulse}. \gls{mls} is a pseudo-random binary sequence that is generated using maximal linear-feedback shift registers. \gls{mls} signals are periodic, have a flat frequency spectrum, and a circular autocorrelation equal to a Kronecker delta function \cite{jacobsen2013fundamentals}, which is the discretized version of the Dirac delta function, making it a suitable excitation signal for \gls{rir} measurements when the \gls{mls} period is longer than the reverberation time of the room \cite{schroeder1979integrated}. The correlation between the \gls{mls} signal and the response signal can be efficiently computed in the frequency domain using the \gls{fwht} algorithm \cite{borish1983efficient, vorlander1997practical}. The main advantage of \gls{mls} is the high \gls{snr} and robustness to impulsive and non-stationary noise sources \cite{nielsen1997improvement,novak2016impulse}. However, the main disadvantage of \gls{mls} signals is the significant non-linear distortion that occurs when the signal is played back through a loudspeaker system. This distortion is caused by loudspeaker non-linearities, which can affect the accuracy of the \gls{rir} measurements \cite{dunn1993distortion, vorlander1997practical, svensson1999errors}.

To mitigate the distortion caused by the non-linear characteristics of loudspeakers when playing back \gls{mls} signals, sine sweeping excitation signals have been proposed. Sine sweep signals offer a higher \gls{snr} as compared to \gls{mls} signals and a complete rejection of harmonic distortion coming from the loudspeaker system \cite{griesinger1996beyond, farina2000simultaneous, muller2001transfer, stan2002comparison, guidorzi2015impulse}. However, sine sweep measurements are sensitive to impulsive noise sources, which, when present, introduce artifacts in the measured \glspl{rir} \cite{griesinger1996beyond, farina2000simultaneous, muller2001transfer, stan2002comparison, guidorzi2015impulse}. Sweeping sines are generated by modulating a sine wave with a time-varying phase $\varphi(t)$ \cite{berkhout1980new, griesinger1996beyond, farina2000simultaneous, muller2001transfer}, i.e.,

\begin{equation}
    s_{\text{sw}}(t) = \sin(\varphi(t)).
\end{equation}

The derivative of the phase denotes the instantaneous frequency of the sweeping sine signal, $f_\text{i}(t) = \frac{d\varphi(t)}{dt}$, which, for room acoustic measurements, can be either linearly increasing \cite{berkhout1980new, muller2001transfer} or exponentially increasing over time \cite{farina2000simultaneous,farina2007advancements, muller2001transfer}. The latter is known as the \gls{ess} signal, which has been shown to have a higher \gls{snr} in the lower frequency range as compared to linearly increasing sine sweeps \cite{farina2000simultaneous}, a desirable property for acoustic measurements. The length of the sine sweep signal should be chosen such that the risk of impulsive noise being present in the recording is minimized, while ensuring that a desired \gls{snr} is achieved \cite{farina2007advancements, guidorzi2015impulse}. Therefore, the choice of excitation signal for \gls{rir} measurements depends on specific requirements, such as the desired \gls{snr}, the mitigation of impulsive noise sources, and the desired accuracy of the measured \glspl{rir}.

A recently proposed method, called Mosaic, uses across-measurement median filtering to recover an impulsive noise-free sweep signal when the sine sweep signal is recorded multiple times \cite{prawda2024stationary}. The Mosaic method is based on the assumption that the impulsive noise sources are uncorrelated across measurements, and that the \gls{rir} is stationary across measurements. The method has been shown to be effective in removing impulsive noise sources from \gls{rir} measurements, and can be used to improve the accuracy of \gls{rir} measurements contaminated with impulsive noise \cite{prawda2024stationary}.

When information about the noise sources in the room is available, \cite{canfielddalifou2018allpass} proposed a method for designing an all-pass chirp signal that allows for the estimation of the \gls{rir} with a constant \gls{snr} across the frequency range of interest. By designing the all-pass chirp signal such that more energy is present at noisy frequency bands, the \gls{snr} remains constant \cite{canfielddalifou2018allpass}.

In \cite{martellotta2009guidelines}, the authors provide guidelines for the measurement of \glspl{rir} in historical buildings, specifically churches. The aim of the study was to simplify and normalize the measurement process of \glspl{rir} in historical buildings, and to improve the comparability of \gls{rir} measurements in different buildings. The authors discuss the choice of source positions, receiver positions, and measurement equipment, which are important factors that affect the accuracy of the measured \glspl{rir}.

\subsubsection{Sound Field Reconstruction in Inaccessible Spaces}
\label{sec:review2024:room_acoustic_acquisition_inaccessible}
In spaces that are inaccessible, partially accessible, or acoustically altered, acoustic measurements are not feasible or desired. In such cases, a 3D digital model of the room can be created in \gls{cad} software from the room's geometry and surface materials. The geometry of the room can be obtained from architectural drawings, paintings, historical documents \cite{sluyts2021archaeoacoustic}, or from a 3D laser scan in case the room is still partially accessible \cite{cui2019automatic, montelpare2024geometrical}. In order to create a realistic acoustic model of the room, the acoustic properties of the room's surfaces need to be modeled by assigning absorption and scattering coefficients \cite{vorlander2020auralization}. This information can be obtained from tables for common building materials or from free-field measurements \cite{vorlander2020auralization}. However, the accuracy of this data is limited, leading to inaccuracies in the simulated \glspl{rir} \cite{vorlander2020auralization}. For accessible or partially accessible spaces, the surfaces of a 3D model are assigned a rough approximation of the absorption and scattering coefficients. During the calibration phase, these parameters are fine-tuned to match measured acoustic properties of the room to within one \gls{jnd} \cite{vorlander2013computer}.

The \gls{rir} between a source and receiver position can be estimated from the room's geometry and surface material properties using simulation techniques. In literature, these techniques can be divided into two main categories, namely: wave-based methods and geometrical acoustics methods \cite{saarelma2019finite}. Wave-based methods, such as the Finite-Difference Time-Domain method, the Finite-Element method, and the Boundary Element Method, numerically solve the wave equation in the room to estimate the \gls{rir}. These methods provide the most accurate results, but are computationally expensive, limiting their use for large and complex geometries \cite{savioja2015overview}. However, as \gls{gpu}-based implementations of wave-based methods can significantly reduce the computation time required for simulating large room acoustics \cite{savioja2010real, savioja2010use, saarelma2019finite}, these methods are becoming more accessible for simulating room acoustics in large and complex geometries. Geometrical acoustics methods, such as the \gls{ism} and \gls{art}, are based on the assumption that sound waves propagate in straight lines and are reflected off the room's surfaces. In \cite{savioja2015overview}, an extensive overview of geometrical acoustics methods for simulating room acoustics is provided. In the remainder of this section, we will focus on the image-source and ray-tracing methods, which are commonly used for estimating \glspl{rir} in simulated environments.

The \gls{ism} was first proposed in the context of simulating rectangular room acoustics by Allen and Berkley \cite{allen1979image}, and later extended to spaces with arbitrary polyhedral geometry \cite{borish1984extension}. \gls{ism} assumes ideal specular reflections from the room's surfaces, and models the room as an infinite number of image sources, which are virtual sources that are mirrored across the room's surfaces. The \gls{rir} is then estimated as the sum of the contributions from all image sources in the room. \gls{ism} can be used to accurately estimate \glspl{rir} in small to medium-sized rooms with simple geometries \cite{allen1979image}. For rectangular shaped rooms with ideal specular reflections, \gls{ism} is a solution to the wave equation \cite{allen1979image}. However, when modeling highly symmetrical spaces using \gls{ism}, such as perfectly rectangular rooms, care should be taken to avoid sweeping echoes \cite{desena2015modeling} that occur due to the regularity with which sound waves are reflected in the room. This effect is known to occur even in real rooms, such as at the staircase of the El Castillo pyramid \cite{declerq2005theoretical,bilsen2006repetition}. To prevent this unwanted effect from occurring in room acoustics simulations, a small random offset can be introduced to each image source position \cite{desena2015modeling}.

When modeling the acoustics of arbitrary geometries using \gls{ism}, the same principle is applied as for rectangular rooms, but only image sources that are visible from the receiver position are considered in the \gls{rir} estimation \cite{borish1984extension}. Visibility checks need to be performed for each image source in the room, which can be computationally expensive for complex geometries. In \cite{schroder2006realtime}, binary space partitioning trees are introduced as a method to efficiently determine the visibility of image sources in arbitrary geometries. The method is shown to be effective in reducing the computational cost of \gls{ism} simulations for complex geometries. However, the method is limited to static setups, and does not account for moving sound sources or receivers in the room.

Hybrid methods for simulating room acoustics exist which combine different simulation techniques to estimate \glspl{rir} for complex geometries \cite{savioja2015overview}. For example, the \gls{ism} and \gls{art}, which will be further discussed below, was combined by Vorländer in 1989 \cite{vorlander1989simulation}, and this approach is used in the commercially available ODEON software package \cite{naylor1993odeon}. The hybrid method uses \gls{art} to discover the location of image-sources that are visible from the receiver position, as these are the image-sources that contribute to the \gls{rir}. For complex geometries, this leads to a significant reduction in the number of image-sources that need to be considered, as the ratio of non-visible to visible image-sources is typically high \cite{vorlander1989simulation}.

The \gls{ism} is a deterministic approach to simulating room acoustics that employs only specular reflection paths. An alternative method for simulating room acoustics based on a room's geometry, is stochastic ray tracing, a popular method for simulating light particle propagation in computer graphics. In the context of room acoustics, ray tracing is used to simulate the propagation of sound waves in an enclosure. This approach was first described for simulating room acoustics by Krokstad \textit{et al.} in 1968 \cite{krokstad1968calculating, krokstad2015early}. In this method, which we will from now on refer to as \gls{art}, sound rays are assigned an initial energy value and are radiated from a source position in the room. The absorption and scattering coefficients are assigned to all surfaces within the room, thereby determining the energy loss and direction of the rays upon reflection off the room's boundaries. The trajectory of the rays is then traced until they reach a desired receiver position within the room. At this position, the energy of each ray is recorded as a function of time. The \gls{rir} is then estimated as the sum of the energy contributions from all rays that reach the receiver position. The main advantage of \gls{art} over the \gls{ism} is the ability to model sound source characteristics, diffuse reflections, and acoustic scattering in complex geometries \cite{savioja2015overview, vorlander2020auralization}. The accuracy of \gls{art} increases with the number of rays that are employed. However, determining the optimal number of rays to use in a simulation is not trivial as it depends on the room's geometry, the desired accuracy of the \gls{rir}, and the available computational resources \cite{savioja2015overview, vorlander2020auralization}. For complex geometries, the number of rays required to accurately simulate the room's acoustics used to be prohibitively high, which leads to long simulation times. However, with the advent of modern computing hardware, such as \glspl{gpu}, the computational time required for \gls{art} simulations has been significantly reduced \cite{savioja2010use}. Under real-time conditions, \gls{art} can be used at the cost of reduced accuracy by using fewer rays in the simulation \cite{vorlander2020auralization}. In \cite{autio2023iterative}, an iterative \gls{art} method was proposed for simulating room acoustics in real-time. The method uses a reduced number of rays in the simulation and refines the simulation iteratively by adding more rays until a desired accuracy is achieved. Preliminary results show the method to be effective in simulating room acoustics in real-time, with a reduced computational cost as compared to traditional \gls{art} methods \cite{autio2023iterative}.

As previously stated, \gls{art} requires a high number of individual rays to accurately simulate the acoustics of a space. To reduce the computational cost further and to increase the efficiency of \gls{art} simulations, volumetric rays, also called beam tracing, can be used \cite{haviland1973monte,savioja2015overview}. The main principle of beam tracing is to expand the rays into volumetric beams that are traced through the room until they reach the receiver position. The beams are then split into smaller beams when they encounter two or more surfaces in the room, which allows for a more efficient simulation of the room's acoustics as compared to traditional \gls{art} methods. For an overview of various ray and beam tracing methods and their application to room acoustics, we refer the reader to \cite{savioja2015overview}. Additionally, a comparison of acoustic simulation and measurement methods for \gls{hws} can be found in \cite{segura2011comparison}.

Commercially available software to simulate room acoustics using \gls{ism} and \gls{art} include Epidaure \cite{vian1992binaural, vanmaercke1993prediction}, ODEON \cite{naylor1993odeon}, CATT-Acoustic \cite{cattacoustic}, Ramsete \cite{ramsete}, EASE \cite{ease}, and Treble \cite{treble}. These software packages are widely used in the field of room acoustics for simulating room acoustics in complex geometries, and are used in the design of concert halls, theaters, and other performance spaces. 

A notable recent development in room acoustics estimation is the use of machine learning and deep learning methods to predict \glspl{rir} \cite{amengual2022room, kelley2024rir,lin2024deep}. These methods rely on large datasets of measured \glspl{rir} to train models that predict the \gls{rir} from a room's geometry and surface material properties. An advantage of machine learning methods is that they can be used to predict additional room acoustics parameters directly from the room's geometry \cite{kim2020real} or captured audio \cite{tang2020scene}. Additionally, the prediction of \glspl{rir} and sound fields by means of generative artificial intelligence methods is an emerging field, in which the two most promising models are \glspl{gan}, e.g. as used in Mesh2IR \cite{ratnarajah2022mesh2ir}, and \glspl{naf} \cite{luo2022learning}. However, the accuracy of machine-learning methods is limited by the quality and quantity of the available training data. To the best of our knowledge, these methods have not been applied yet to the prediction of \gls{hws} acoustics.

\subsection{Acoustic acquisition of Historical Worship Spaces}

\noindent In \gls{hws}, the acoustics of the space are of particular interest due to the unique architectural design of the spaces, which can have a significant impact on the sound field within the space. In spaces that remain unaltered and accessible, acoustic measurements can be conducted to preserve and analyze the acoustics of the space \cite{navarro2009western, malecki2017acoustics}. Before current methods for measuring \glspl{rir} were developed, the reverberation time was measured by exciting the space with 1/6 octave band filtered white noise signals, and measuring the acoustic decay pattern \cite{raes1953measurements, anderson2000acoustic}. A study from 1971 measured the reverberation time of two Basilicas in Rome at single frequencies using an organ pipe as a sound source \cite{shankland1971acoustics}. In \cite{sügül2013impact}, \gls{rir} measurements of a mosque in Ankara, Turkey, were conducted by exciting the space with broadband white noise.

The \gls{mls} method has been used to measure the acoustics of various \gls{hws}, including Catholic churches in Poland \cite{kosala2013assessing, kosala2016calculation}, 21 mosques in Saudi Arabia \cite{abdou2003measurement}, nine Romanesque churches in Southern Italy \cite{cirillo2002acoustics, cirillo2003acoustics}, 12 Mudejar-Gothic churches in Seville, Spain \cite{galindo2008correlations, galindo2009acoustic}, six churches in Switzerland and Portugal \cite{desarnaulds2002church}, and the Cathedral of Benevento in Italy \cite{iannace2016acoustic}.

Sine sweep methods, which offer a higher \gls{snr} and complete rejection of harmonic distortion from the loudspeaker system, have been used extensively to measure the acoustics at various positions in \gls{hws}. The method has been used to measure the acoustics in churches \cite{devries2002measurements, tronchin2008acoustical, martellotta2008subjective, martellotta2009identifying, soeta2012effects, alonso2016room, martellotta2016understanding, alvarezmorales2017virtual, dorazio2020understanding, tronchin2020evaluation, alberdi2021acoustic, ricciutelli2024churches}, cathedrals \cite{alvarez2011virtual, alonso2014acoustic, alonso2017virtual, martellotta2018investigation, katz2020acoustic, alvarezmorales2020acoustic}, Buddhist temples \cite{soeta2013measurement}, Egyptian temples \cite{warusfel2021assessing}, mosques \cite{fausti2003comparing, suarez2018virtual, sugul2021exploration, kitapci2021acoustic}, and historic monuments such as Stonehenge \cite{fazenda2013recreating, cox2020scale}.

Alternative methods, such as balloon bursts and starter pistol shots, have been used less frequently due to the lack of repeatability and the potential for inaccuracies in the measured \glspl{rir}. However, in spaces with no access to mains power, the use of balloon bursts or starter pistol shots may be the only viable option for measuring the acoustics of \gls{hws} \cite{tzekakis1975reverberation, pyshin2001acoustics, patynen2011investigations, pedrero2014acoustical, katz2020acoustic, elkhateeb2021acoustics, sukaj2021two, sukaj2022byzantine}.

When the acoustics of \gls{hws} have been altered over time, when the space is no longer accessible, or when a complete acoustic reconstruction of the space is desired, the methods described in Section \ref{sec:review2024:room_acoustic_acquisition_inaccessible} can be used to reconstruct the acoustic sound field of the space from a geometrical model of the room and its surface material properties. This has been done for various \gls{hws} including mosques \cite{weitze2003comparison, orfali2005acoustical, sügül2013impact, sugul2016investigations, kavraz2016acoustic, suarez2018virtual, kitapci2021acoustic, sugul2021exploration}, ancient Greek spaces \cite{koutsivitis2005reproduction, bellia2023experiencing}, churches \cite{galindo2009acoustic, suarez2013acoustics, berardi2013simulation, alvarez2015geometrical, suarez2016archaeoacoustics, alonso2016room, zhao2016preliminary, sender2018virtual, hudokova2021impact, dorazio2020understanding, tronchin2020evaluation, alberi2021evolutionary, alberdi2021acoustic, autio2021historically, ricciutelli2024churches, alambeigi2024auralising}, cathedrals \cite{alvarez2011virtual, alonso2014acoustic, suarez2015intangible, postma2016virtual, iannace2016acoustic, alvarezmorales2017virtual, alonso2017virtual, martellotta2018onsite, eley2021virtual, alvarezmorales2020acoustic, mullins2022development, canfielddafilou2022opening, canfielddafilou2023can, mullins2023immersive, canfielddafilou2024voices, demuynke2024ears}, and Buddhist temples \cite{soeta2013measurement}.

Due to imperfections in the modelled room geometry and surface material properties, the geometrical model only approximates the true space, limiting the accuracy of the simulated \gls{rir}. When the \gls{hws} is still or partially accessible, the accuracy of the simulated \glspl{rir} can be improved by calibrating the model to measured \glspl{rir} from the space \cite{alvarez2011virtual, soeta2013measurement, alonso2014acoustic, postma2015creation, postma2015calibrated, alvarez2015geometrical, postma2016perceptive, postma2016virtual, iannace2016acoustic, alonso2016room, alvarezmorales2017virtual, alonso2017virtual, martellotta2018onsite, suarez2018virtual, dorazio2020understanding, alvarezmorales2020acoustic, kitapci2021acoustic, sugul2021exploration, alberi2021evolutionary, eley2021virtual, autio2021historically, ricciutelli2024churches, canfielddafilou2024voices, demuynke2024ears}.

In \cite{segura2011comparison}, a comparison of various room simulation and measurements techniques for \gls{hws} was conducted. The study compared the acoustic parameters of a Basilica in Valencia, Spain, measured using \gls{mls} and sine sweep methods, with the simulated acoustics of the space using simulation software CATT-acoustic and Epidaure.

A methodology for studying the acoustic environment of cathedrals was presented in \cite{alvarezmorales2014methodology}. The authors used both experimental measurements and computer simulations to analyze the cathedral of Malaga as a case study. The acoustic behavior of the cathedral at relevant positions for liturgical activities, including the altar, choir, pulpit, organ, and retrochoir, and for different listener zones within the cathedral were investigated. This study aimed to answer how the sound field is influenced by architectural design of the cathedral and how it impacted various liturgical and cultural activities, including music and speech intelligibility \cite{alvarezmorales2014methodology}.

It should be noted that acoustic measurements of historically valuable spaces other than worship spaces have been carried out in the context of archaeoacoustics research at various locations around the world \cite{navas2023archaeoacoustics}.

Commercially available software packages for room acoustic simulation and their use in simulating the acoustics of \gls{hws} are listed in Table \ref{tab:review2024:acoustic_sim_software}. The software packages are widely used in the field of room acoustics for simulating room acoustics in complex geometries, and are used in the design of concert halls, theaters, and other performance spaces.

\begin{table}[h!]
    \centering
    \caption{Available software packages for room acoustic simulation and their use in \gls{hws}.}
    \begin{NiceTabular}{X[l, 3] X[l, 2] X[l, 3]}
        \toprule
        Software & License & Use in \gls{hws} \\
        \midrule
        ODEON \cite{naylor1993odeon} & Commercial & \cite{weitze2003comparison, alvarezmorales2020acoustic, autio2021historically, sügül2013impact, iannace2016acoustic, dorazio2020understanding, ricciutelli2024churches, soeta2013measurement, sugul2021exploration, kitapci2021acoustic, kavraz2016acoustic, bellia2023experiencing, hudokova2021impact,  alambeigi2024auralising} \\
        CATT-Acoustic \cite{cattacoustic} & Commercial & \cite{postma2015creation,postma2016virtual,postma2016perceptive,alvarezmorales2017virtual,suarez2018virtual,eley2021virtual,mullins2023immersive,segura2011comparison,galindo2009acoustic,alonso2016room,alberdi2021acoustic,alvarez2011virtual,alonso2014acoustic,alonso2017virtual,suarez2013acoustics,berardi2013simulation,alvarez2015geometrical,suarez2016archaeoacoustics,sender2018virtual,alberi2021evolutionary,suarez2015intangible,martellotta2018onsite,mullins2022development,canfielddafilou2022opening,canfielddafilou2023can,canfielddafilou2024voices,demuynke2024ears,postma2015calibrated,alvarezmorales2014methodology} \\
        Epidaure \cite{vian1992binaural, vanmaercke1993prediction} & Commercial & \cite{segura2011comparison} \\
        Ramsete \cite{ramsete} & Commercial & \cite{tronchin2020evaluation} \\
        EASE \cite{ease} & Commercial & \cite{sugul2016investigations}\\
        Treble \cite{treble} & Commercial & \\
\bottomrule
    \end{NiceTabular}
    \label{tab:review2024:acoustic_sim_software}
\end{table} \section{Sound Field Analysis}
\label{sec:review2024:sound_field_analysis}

\noindent As discussed in the previous section, the \gls{rir} fully describes the acoustic wave propagation between a sound source and receiver in an enclosed space. However, the \gls{rir} does not provide a spatial representation of the larger sound field in the room. The sound field for a given sound source positioned in the space is the three-dimensional representation of sound pressure and pressure velocity around a sensor position in the room. For simulated room acoustics, the analysis of the sound field is trivial, as the entire sound field can be computed at every point in space. For real-world measurements however, the sound field is not directly accessible and must be inferred from acoustic measurements. The sound field in a space can be analyzed from microphone array \gls{rir} measurements, which carry the spatial information of the sound field. \gls{sfa} aims to extract the spatio-temporal characteristics of sound propagation around a sensor array from measurements made in the sound field. This spatial information can be used for room acoustic analysis and design, validation of room acoustic simulations, as well as for auralization of the room acoustics.

\subsection{Methods}

\noindent \gls{sfa} is the process of extracting spatial information from the sound field in a space using microphone array measurements. The geometry of the microphone array, the number of microphones, and the microphone spacing are important factors that affect the spatial resolution of the analysis. Depending on the geometry and properties of the microphone array, different spatial analysis methods are used. The most common microphone array geometries for \gls{sfa} are linear arrays, circular arrays, and spherical arrays. The former two are used for two-dimensional \gls{sfa}, while the latter can be used for three-dimensional \gls{sfa}.

One of the earliest methods that enabled the analysis of sound fields in an enclosed space using a linear microphone array is \gls{wfa}. This analysis method was presented by Berkhout \textit{et al.} in 1997 \cite{berkhout1997array}. The method is directly related to \gls{wfs}, a sound field synthesis method that will be further discussed in Section \ref{sec:review2024:sound_field_synthesis}. \gls{wfa} is a spatial analysis method that provides insight into the temporal and spatial structure of the sound field by transforming linear microphone array measurements from the space-time domain to the so-called "ray parameter versus intercept time" domain using the Linear Radon Transform. This transformation decomposes the measured data into a set of traveling plane wave components, which can be used to analyze properties of a sound field, such as diffusivity, lateral energy content, and allows for the discrimination and analysis of reflection and diffraction events in the sound field \cite{baan1998array, devries1999wave, kuster2004acoustic}. In \cite{berkhout2007array}, \gls{wfa} was adapted to analyze and simulate bending wave fields of a vibrating steel plate using linear sensor arrays. 

\begin{figure}
    \centering
    \includegraphics[width=\columnwidth]{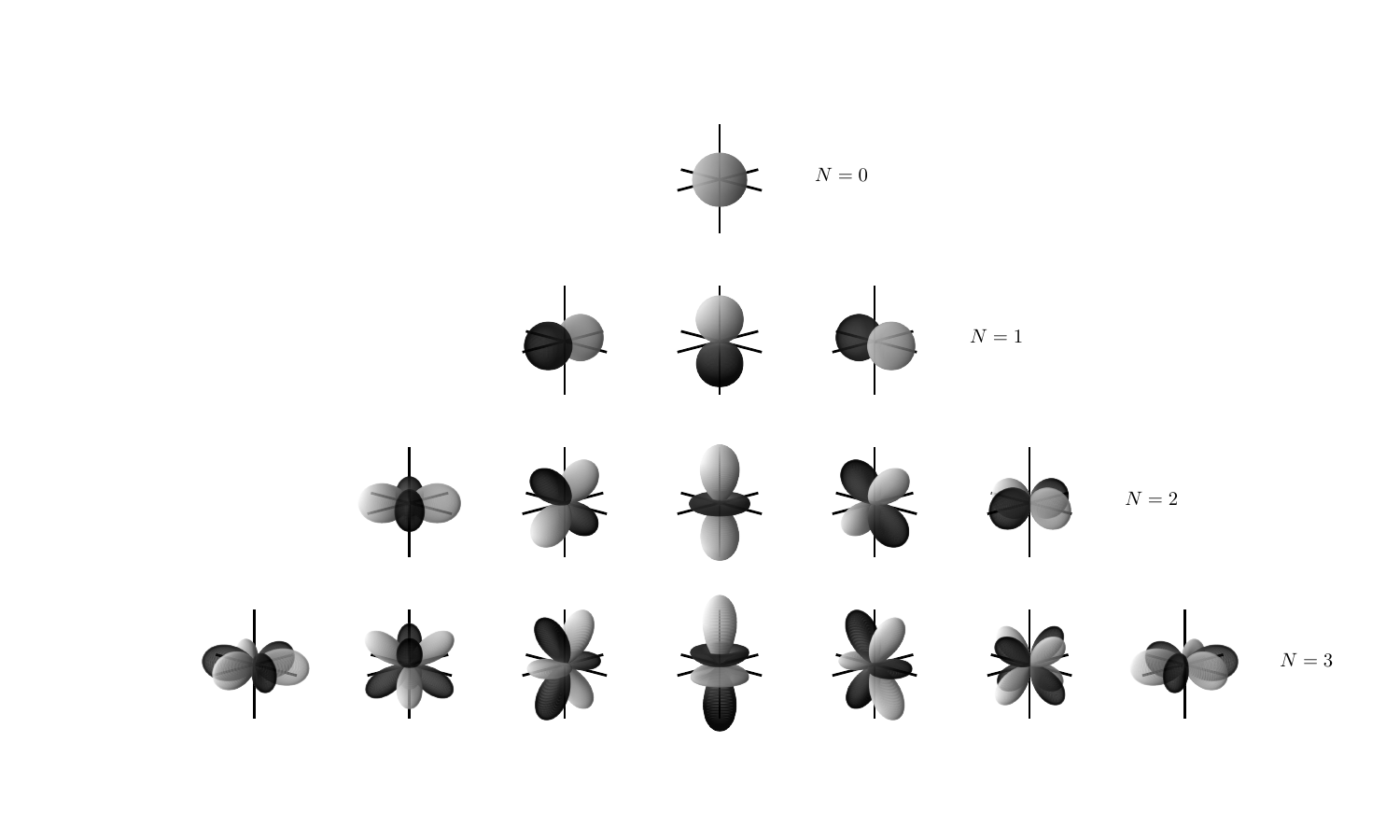}
    \vspace{-4em}
    \caption{Magnitude of spherical harmonics up to order $N=3$. Light regions represent positive sign values, while dark regions represent negative sign values. The radius indicates the magnitude of the spherical harmonic.}
    \label{fig:review2024:spherical_harmonics}
\end{figure}

An alternative method for analyzing the sound field on a circular or spherical surface within an enclosed space is the \gls{shd} method. The method can be used with circular or spherical microphone arrays for two- and three-dimensional \gls{sfa} \cite{abhayapala2010spherical, chen2015sound}. The \gls{shd} method decomposes the sound field captured by the microphone array into spherical harmonic components that form an orthogonal basis on the surface of a sphere. The spherical harmonics can be used to represent the sound field in the room, where the coefficients of the spherical harmonics describe the spatial distribution of the sound field. The \gls{shd} method is a powerful tool for \gls{sfa} and has been used in various applications such as beamforming \cite{sun2010spherical, rafaely2010spherical, huang2016real, kumar2016near}, direction of arrival estimation \cite{tervo2015direction, weijian2016direction}, plane wave decomposition \cite{park2005sound, zotkin2010plane, imran2016plane}, and spatial sound field reproduction through Ambisonics \cite{gerzon1973periphony, gerzon1985ambisonics}. The latter will be further discussed in Section \ref{sec:review2024:sound_field_synthesis}. The order of the spherical harmonics determines the spatial resolution of the analysis, with higher-order spherical harmonics providing higher spatial resolution, as shown in Fig. \ref{fig:review2024:spherical_harmonics} where the magnitude of real spherical harmonics up to order $N=3$ is visualized. However, higher-order spherical harmonics require a larger number of microphones to be accurately captured, specifically $(N+1)^2$ microphones for analyzing up to order $N$ spherical harmonics \cite{rafaely2010spherical}.

\gls{wfa} and \gls{shd} are wave-based analysis methods which aim to analyze the true physical sound field in the room over an area surrounding the microphone array. In contrast, the \gls{sdm} \cite{tervo2013sdm} is a parametric method that models the \gls{rir} between a source and a microphone array as the sum of a finite number of image sources arriving at the microphone array's geometric center \cite{allen1979image}. The method assumes that every sample in the pressure signal recorded at the center of the microphone array, is the result of a single acoustic event, i.e., image source, that can be described by a certain \gls{doa} $\mathbf{\theta}(t)$ and reference pressure value $p(t)$. When the \gls{doa} is known for every sample in the reference pressure signal, the sound field can be reconstructed by summing the contributions of all localized image sources.  

The \gls{sdm} analyzes the multichannel \glspl{rir} captured by the microphone array by applying a sliding short-time windowing function to the measured \glspl{rir}. The windowing function is non-zero only for the interval $t \in [\kappa - \frac{L}{2} \dots \kappa + \frac{L}{2}]T$, where $L$ denotes the window length and $T = 1 / f_s$ is the sampling period. The window length should be chosen to exceed the time it takes for a wavefront to propagate through the microphone array, and is therefore dependent on the microphone spacing. By sliding the short-time window over the \glspl{rir} at every discrete time step $T$, a set of overlapping short-time windows is obtained. The \gls{doa} $\mathbf{\theta}(t)$ can be estimated for every short-time window position by applying a localization algorithm to each short-time window in the set of \glspl{rir}. The localization algorithm can be implemented using the \gls{gcc} \cite{knapp1976generalized,tervo2013sdm} for open microphone arrays, or using pseudo-intensity vectors for B-format microphone arrays \cite{jarrett2010source, zaunschirm2018brir, amengual2020optimizations}. At low sampling rates, the localization performance can be improved using interpolation techniques \cite{lai1999interpolation,zhang2005cross, rosseel2021improved}. The reference pressure signal $p(t)$ is measured using an omnidirectional microphone located at the center of the microphone array. In case there is no microphone located in the geometric center of the array, the reference pressure signal can be predicted from the image-source locations \cite{tervo2013sdm}. The estimated \glspl{doa} and reference pressure values can be used to visualize the directionality of the sound field in the room. This has been shown to be useful for the analysis of sound fields in large spaces, specifically for the analysis of concert halls \cite{patynen2013spatial, tervo2013spatio, amengual2016physical, moller2023spatial}, nightclubs \cite{tervo2015spatial}, music studios \cite{tervo2014preferences}, car cabins \cite{tervo2015spatialcar}, and chamber music hall acoustics \cite{lachenmayr2023chamber}.

While the \gls{sdm} is a powerful method for the analysis of sound fields in enclosed spaces, it has some limitations \cite{meyerkahlen2022what}. The main assumption of the \gls{sdm} is that for every overlapping analysis window, precisely one acoustic reflection, i.e., image source is present. This assumption is not always valid in practice, i.e., when the microphone array is located close to a reflective surface, when the room's geometry is highly symmetrical or when considering the late part of the \gls{rir} where the reflection density is high. In these cases, the \gls{sdm} may not accurately estimate the \gls{doa}. In these cases, the analysis window length should be kept to a minimum to reduce the number of overlapping reflections in the analysis window.

\subsection{Spatial sound field analysis of historical worship space acoustics}

\noindent It is known that the spatial properties of the sound field in \gls{hws} vary significantly across the space due to the complex geometry and surface material properties of these spaces. The use of \gls{sfa} in \gls{hws} allows researchers to analyze the unique spatial properties at different locations within the space, which can be used to better understand the acoustics at choir stalls, organ lofts, and other important locations within the space. The spatial analysis of the sound field in \gls{hws} can also be used to evaluate the impact of architectural changes, such as the deployment of sound-absorbing materials, on the spatial distribution of sound energy in the space.

In 2002, de Vries \textit{et al.} used a circular microphone array for \gls{rir} measurements inside a church in Weimar, Germany, to analyze the spatial sound field for the reproduction of choir concert recordings using \gls{wfs} \cite{devries2002measurements}. The study found that the circular microphone array provided a good spatial resolution of the sound field in the church, which was essential for the accurate reproduction of the choir concert recordings using \gls{wfs}.

Spatial \gls{sfa} based on \gls{shd} was used to analyze and visualize the spatial energy distribution of the measured sound field in churches \cite{elicio2015acoustics, alary2019assessing}, cathedrals \cite{martellotta2018investigation, alvarezmorales2020acoustic}, and an ancient Egyptian temple \cite{warusfel2021assessing}

The spatial energy distribution of a church in Lyddington, UK, was analyzed using the \gls{sdm} in \cite{frank2016spatial}. In 2017, Amengual \textit{et al.} used a tetrahedral microphone array to analyze the spatial sound field of the reproduced acoustics of a hall, studio, and church \cite{amengual2017spatial} using the \gls{sdm}. The study compared the spatial resolution of the \gls{sdm} with IRIS \cite{iris2024}, a commercial \gls{sfa} software, and found that both methods provided similar spatial resolution in simple sound fields. However, in complex and dense sound fields, such as those found in \gls{hws}, the higher temporal resolution of the \gls{sdm} provided a better spatial energy distribution over IRIS. In 2020, Katz and Weber \cite{katz2020acoustic} used the \gls{sdm} to analyze the energy distribution of the sound field in the Cath{\'e}drale Notre-Dame de Paris in pre- and post-fire conditions. \section{Sound Field Synthesis}
\label{sec:review2024:sound_field_synthesis}

\noindent The spatial analysis of the sound field in a room provides valuable information about the spatial distribution of sound pressure and pressure velocity around a sensor position in the room. This information can be used for auralization of room acoustics, as well as for room acoustic analysis and design, as shown in the previous section. \gls{sfs} is the process of auralizing, i.e., recreating, the sound field in a room such that the listener perceives the sound field as if they were present in the measured space. In this section, we will discuss various \gls{sfs} methods, including loudspeaker synthesis and headphone synthesis, and how the spatial information obtained from \gls{sfa} can be used for \gls{sfs}.

\subsection{Methods}

\subsubsection{Loudspeaker Synthesis}

\noindent Loudspeaker synthesis is the process of recreating a sound field of a room using an array of loudspeakers. The loudspeakers are positioned in the listening room such that the sound field is either synthesized at a specific listening position, referred to as the \textit{sweet spot}, or over an extensive listening area which allows for listeners to experience the synthesized sound field as they move around the room. The number of loudspeakers and their positioning in the room are important factors that affect the spatial resolution of the synthesis. While the loudspeaker configuration can be arbitrary, common loudspeaker configurations for \gls{sfs} are linear arrays, circular arrays, and spherical arrays. The former two are used for two-dimensional \gls{sfs}, while the latter can be used for three-dimensional \gls{sfs}.

Loudspeaker synthesis is achieved by constructing synthesis filters from the spatial information of the sound field. The synthesis filters are then convolved with a dry audio signal, such as speech or music to create the synthesized sound field.

A sound field consists of a set of auditory events, see \eqref{eq:review2024:rir_events}, also called virtual sources, that are perceived as coming from a certain direction. Accurate reproduction of sound fields using a set of loudspeakers creates the illusion that the virtual source is coming from a specific direction, regardless of the loudspeaker positioning in the reproduction setup. A common method to achieve this is by mapping the virtual sources to the loudspeaker array in a way that the sound source is perceived as coming from a specific direction. A straightforward method, called \gls{nls} achieves this by mapping each virtual source to the loudspeaker that is closest to the desired direction of the virtual source \cite{tervo2015spatialcar}. This method works well for loudspeaker arrays containing many loudspeakers, but is not suitable for loudspeaker arrays with few loudspeakers as the spatial resolution of the synthesis is limited by the loudspeaker spacing. To achieve higher spatial resolution in the synthesis, the virtual source can be positioned between loudspeakers in the array. This can be achieved using amplitude and/or time-delay differences between the loudspeakers, a technique known as panning \cite{desena2015efficient}. Panning is a common technique used in audio production to position a sound source in a stereo or multichannel loudspeaker setup. The simplest form of panning in a stereo loudspeaker setup is amplitude panning, where the amplitude of the sound source is varied between the left and right loudspeakers to create the illusion that the sound source is coming from a specific direction. In a multichannel loudspeaker setup, amplitude panning can be extended to \gls{vbap} \cite{pulkki1997virtual, pulkki2001localization} which is a generalized amplitude panning technique for positioning a sound source between apertures of loudspeaker triplets. In \gls{vbap}, the loudspeaker directions for a loudspeaker triplet are represented by three-dimensional unit vectors $\mathbf{v}_1$, $\mathbf{v}_2$, and $\mathbf{v}_3$, and the desired sound source direction is represented by a three-dimensional vector $\mathbf{v}_s$. The sound source is then panned to the loudspeaker triplet by calculating the gains $g_1$, $g_2$, and $g_3$ for each loudspeaker in the triplet using the following equation \cite{pulkki1997virtual},

\begin{equation}
    [g_1, g_2, g_3] = \mathbf{v}_s^T \mathbf{V}^{-1},
\end{equation}

\noindent where $\mathbf{V} = [\mathbf{v}_1, \mathbf{v}_2, \mathbf{v}_3]^T$ is the matrix of loudspeaker directions. The gains $g_1$, $g_2$, and $g_3$ are then used to drive the loudspeakers in the triplet. \gls{vbap} is a flexible and efficient method for panning sound sources to arbitrary loudspeaker layouts. In addition, the loudspeaker gains are normalized to ensure that the sound source is not amplified or attenuated during panning \cite{pulkki1997virtual, pulkki2001localization}. A shortcoming of intensity panning in \gls{vbap} is the nonuniform spatial spread of the panned virtual source \cite{pulkki1999uniform}. To address this, \gls{mdap} was proposed by Pulkki in 1999 \cite{pulkki1999uniform}, which is a method that uses multiple loudspeaker triplets to pan a sound source to a specific direction. The method ensures that the panned virtual source is uniformly spread over the loudspeaker array \cite{pulkki1999uniform}.

Panning techniques can be used to position virtual sources around a listener in a room using loudspeaker arrays. In Section \ref{sec:review2024:sound_field_analysis}, the \gls{sdm} was discussed as a parametric method to analyze the sound field in a room using microphone array measurements by decomposing the sound field into a set of image sources described by their \gls{doa} $\mathbf{\theta}(t)$ and reference pressure value $p(t)$. From this, the sound field can be reproduced by positioning the virtual sources around the listener using panning techniques such as \gls{vbap} or \gls{nls} \cite{tervo2013sdm, tervo2015spatialcar}.

In \cite{patynen2014amplitude}, it was shown that the \gls{sdm} in combination with amplitude-based panning techniques can decrease the perceived brightness of the reproduced sound field. Therefore, a trade-off between the perceived brightness of the sound field and the spatial resolution of the synthesis must be considered when using panning techniques for \gls{sfs}.

In 1988, Berkhout introduced the concept of \gls{wfs} \cite{berkhout1988holographic}, a holographic \gls{sfs} method which reproduces sound fields that are perceptually indistinguishable from the measured sound field. \gls{wfs} was introduced as an auralization method for room acoustics in 2004 by de Vries and Hulsebos \cite{devries2004auralization}. The basis of \gls{wfs} can be found in Huygens' principle, which states that a wavefront can be considered as a continuous distribution of secondary spherical wave fronts. The wavefront can be synthesized during reproduction by summing the contributions of all secondary sources \cite{berkhout1988holographic, berkhout1993acoustic} on the boundary surface. Mathematically, \gls{wfs} is based on the Kirchhoff-Helmholtz integral, which states that, when the sound field on the boundary surface $S$ of a closed, source-free volume $V$, is known in terms of pressure and normal particle velocity, the sound pressure can be calculated at any point in the volume $V$ \cite{berkhout1988holographic, berkhout1993acoustic}. The sound field information on the boundary surface $S$ can be obtained by measuring the sound field using microphone arrays, similarly to \gls{wfa}. In practice, due to a finite number of microphones and loudspeakers that can be used during the analysis and reproduction, the distribution of secondary sources on the surface is not continuous, but discrete. This discretization of the boundary surface introduces spatial aliasing and truncation artifacts in the synthesized sound field. The spatial aliasing artifacts are caused by the finite loudspeaker spacing, which limits the frequency range that can be accurately reproduced by \gls{wfs}, known as the spatial aliasing frequency \cite{verheijen1997sound}. The spatial aliasing frequency is dependent on the loudspeaker spacing $\Delta x$ and the speed of sound $c$ in the medium, and is given by,

\begin{equation}
    f_\textrm{max} = \frac{c}{2\Delta x}.
\end{equation}

\noindent To avoid spatial aliasing, the loudspeaker spacing must be smaller than half the wavelength of the highest frequency in the sound field that is to be reproduced. The truncation artifacts are caused by the finite size of the loudspeaker array. A discussion and analysis of the spatial aliasing and truncation artifacts in \gls{wfs} can be found in \cite{spors2006spatial, ahrens2010comparison, gergely2018spatial, winter2019geometric}.

In \cite{hulsebos2002improved}, three two-dimensional microphone array configurations were considered for \gls{wfs}, namely: a linear array, a cross array, and a circular array. This study found that the circular array configuration provided the best performance in terms of homogeneous spatial resolution and artifact-free reconstruction area for auralization using \gls{wfs} \cite{hulsebos2002improved}. While circular and three-dimensional microphone arrays can be used for \gls{wfa}, no literature was found that specifically addresses the use of \gls{wfs} with three-dimensional microphone arrays. A method to extract three-dimensional sound field information from two-dimensional microphone array measurements using \gls{wfs} was, however, proposed by \cite{devries2007extraction}.

Recently, Amengual Gar{\'i} presented an exploratory study that combined the \gls{sdm} analysis and \gls{wfs} for \gls{sfs} \cite{amengual2024spatial}. The study used the \gls{sdm} to decompose the sound field in a room into a set of image-sources that were reproduced using \gls{wfs} by modeling the image-sources as either point sources or plane waves. When modeling the image-sources as plane waves, the synthesized sound field showed good agreement with the measured sound field. However, further research is still needed to evaluate the perceptual performance of the combined \gls{sdm} and \gls{wfs} method for \gls{sfs} \cite{amengual2024spatial}.

In Section \ref{sec:review2024:sound_field_analysis}, the \gls{shd} method was discussed as a spatial analysis method to analyze the sound field in a room by representing the sound field as a sum of spherical harmonic components. The spatial representation that is retrieved from \gls{shd} can be used for \gls{sfs} using Ambisonic or \gls{hoa} synthesis \cite{gerzon1973periphony, gerzon1985ambisonics}. By decoding the spherical harmonic components of the sound field to a loudspeaker array using Ambisonic decoding equations \cite{gerzon1977multi1,gerzon1977multi2}, the sound field can be reconstructed. For first-order Ambisonic reproduction, a well-known method for decoding the spherical harmonic components to arbitrary and irregular loudspeaker layouts, was presented by Gerzon in 1992 and referred to as Vienna decoders \cite{gerzon1992ambisonic}. For higher-order Ambisonic decoding, \gls{allrad} was presented by Zotter and Frank in 2012 \cite{zotter2012allround}. This method combines Ambisonic decoding with \gls{vbap} to provide a flexible and efficient method for decoding higher-order Ambisonic signals to arbitrary loudspeaker layouts by first decoding the spherical harmonic components to a virtual loudspeaker array, after which the virtual loudspeaker signals are panned to the physical loudspeaker array using \gls{vbap}.

A study by Frank and Zotter \cite{frank2016spatial} in 2016 investigated the dependency of the spatial impression of synthesized reverberant sound fields on the directional resolution using \gls{hoa} synthesis \cite{frank2016spatial}. The original sound field was measured using a first-order Ambisonic microphone array, and the \gls{sdm} was used to create first-, third-, and fifth-order Ambisonic representations of the sound field. The Ambisonic representations were then decoded to a loudspeaker array and the synthesized sound field was presented to listeners in a listening room. The study found that auralizations using higher-order Ambisonic representations increased the perceptual sweet spot of the synthesized sound field, and provided well-focused direct sound \cite{frank2016spatial}.

Ambisonic and \gls{wfs} are both holophonic, wave-based sound field reproduction methods, meaning that they aim to physically reconstruct the measured sound field in a room. While the two methods are based on different principles, it has been shown that \gls{wfs} and \gls{hoa} are equivalent under certain conditions as they are both able to provide an exact solution to the wave equation \cite{rozenn1999sound, daniel2003further}.

\subsubsection{Headphone Synthesis}

\noindent Headphone synthesis, also known as binaural synthesis, is the process of recreating the sound field of a room for an individual listener by means of headphones. The main advantage of using headphones for \gls{sfs} is that the reproduced and listener's acoustic environment can be decoupled, which allows for a more controlled listening experience. However, headphone synthesis also presents some challenges, such as the feeling of being closed off from the environment, and in-head localization, which is the perception of sound sources coming from inside the listener's head, leading to unnatural listening experiences \cite{vorlander2020auralization}. To prevent in-head localization and allow for spatialization, the sound field must be synthesized by taking into account the listener's individual \glspl{hrtf}. The \gls{hrtf} is the direction-dependent transfer function describing the filtering effect of the listener's head, shoulders, torso, and pinnae on a sound wave traveling from free field to the listener's eardrum \cite{wenzel1993localization}. The \gls{hrtf} is unique to each listener, and therefore, personalized \glspl{hrtf} are required for accurate headphone synthesis.

In practice, obtaining personalized \glspl{hrtf} is done by placing a probe microphone in the listener's ear canal and measuring the acoustic transfer function from a sound source placed in the free field to the probe microphone. This process is repeated for both ears and for a wide range of sound source directions. While these measurements provide the most accurate \glspl{hrtf} for a listener, the process is costly and time-intensive \cite{vorlander2020auralization}. Therefore, publicly available \glspl{hrtf} are often used for headphone synthesis. Commonly used \gls{hrtf} databases are the \textit{CIPIC} database, which contains individual HRTFs of 45 subjects \cite{algazi2001cipic}, and the \textit{ARI} database, which contains HRTFs of more than 250 subjects \cite{ari2024hrtfs}. For a detailed comparison of different HRTF databases, see \cite{iida2019head}.

In real acoustic environments, the listener's head orientation can change freely, influencing the directional perception of the sound sources in the acoustic environment. To replicate this realism, and enhance the listening experience, the listener's head orientation must be tracked during binaural synthesis. This tracking allows virtual sound sources to adjust their positions relative to the listener's head, ensuring that the spatial audio experience remains consistent and immersive, even as the listener moves their head.  This dynamic adjustment effectively fixes the virtual sources around the listener's head, maintaining their perceived locations regardless of the listener's head orientation.

Head tracking in 6-\gls{dof} can be achieved through various methods, including electromagnetic, camera-based, and optical tracking \cite{laitinen2012influence, amengual2019flexible}, as well as through the use of \glspl{imu} that integrate inertial sensors such as accelerometers, gyroscopes, and magnetometers, which are mounted on the listener's head or headphones \cite{hendrickx2017influence, otani2020binaural, francek2023performance}. For a comprehensive review of head tracking techniques for binaural synthesis, the reader is referred to \cite{hess2012head}. In \cite{hendrickx2017influence}, it was shown that head tracking for binaural synthesis improved the externalization of virtual sources when using non-individualized \glspl{hrtf}. Moreover, head tracking reduced the number of front-back confusions in the localization of virtual sources, which is a common issue when using non-individualized \glspl{hrtf} \cite{wightman1999resolution}.

Binaural synthesis is achieved by mapping the spatial information of the measured sound field to a set of virtual loudspeakers positioned around the listener's head. Each virtual loudspeaker corresponds with a specific \gls{hrtf} depending on the relative direction of the virtual loudspeaker to the listener's head. When the listener moves their head, the \glspl{hrtf} are adjusted accordingly to maintain the illusion that the virtual loudspeaker array is fixed around the listener's head as the \glspl{hrtf} provide the frequency response characteristics necessary to recreate the correct spatial audio cues, enabling the perception of sound originating from distinct directions relative to the listener's head. The binaural synthesis filters, one for each ear, are created by convolving each virtual loudspeaker, encoded by the spatial information of the measured sound field, with the corresponding \gls{hrtf}. These convolutions are then superimposed to create binaural synthesis filters that can be used for auralization of a dry audio signal by convolving this signal with the binaural synthesis filters and presenting the resulting binaural signals to the listener through headphones. The \gls{sfs} methods used for loudspeaker synthesis can be adapted to binaural synthesis by mapping the spatial information of the sound field to these virtual loudspeakers, instead of physical loudspeakers.

The spatial information obtained from the \gls{sdm} analysis, i.e., the \gls{doa} $\mathbf{\theta}(t)$ of the image sources and reference pressure value $p(t)$, can be used for binaural synthesis by panning the image sources to the virtual loudspeakers using \gls{vbap} or \gls{nls} \cite{patynen2014amplitude, puomio2017optimization, zaunschirm2018brir}. Since the virtual loudspeakers can be positioned in a dense grid around the listener's head, usually with a spatial resolution of $2.5^\circ-5^\circ$ \cite{iida2019head}, the achievable spatial resolution of binaural synthesis is significantly higher compared to loudspeaker synthesis, which motivates the use of \gls{nls} for binaural synthesis over \gls{vbap} \cite{patynen2014amplitude}. In \cite{amengual2020optimizations}, several improvements to the \gls{sdm} method for binaural synthesis were proposed, including the use of a spherical microphone array for \gls{sfa}, the optimal microphone array size and short-time window length for \gls{doa} estimation, and an equalization method to reduce spectral whitening in the late reverberation due to rapidly varying \gls{doa} estimates. The improvements were evaluated with perceptual experiments. In \cite{ahrens2019perceptual}, the perceptual performance of the \gls{sdm} for binaural synthesis was evaluated using a variety of microphone array geometries. One perceptual study showed that binaural synthesis using the \gls{sdm} is indistinguishable from loudspeaker synthesis in the absence of visual cues \cite{amengual2019flexible}. However, when visual cues were present, listeners could distinguish between binaural and loudspeaker synthesis, indicating that visual cues play a significant role in the perception of \gls{sfs} methods \cite{amengual2019flexible}.

Binaural \gls{wfs} can be achieved by creating a set of virtual loudspeakers around the listener's head and using the \gls{wfs} method to synthesize the sound field to these virtual loudspeakers \cite{wierstorf2013binaural}. In a binaural listening experiment, Wierstorf \textit{et al.} showed that acoustic reproduction using \gls{wfs} suffers from coloration artifacts due to spatial aliasing occurring when the inter-loudspeaker distance is larger than the wavelength of the sound field \cite{wierstorf2014coloration}. The coloration can be reduced by using \glspl{hrtf} with a higher spatial resolution, which effectively decreases the inter-loudspeaker distance in the binaural synthesis.

For binaural Ambisonic and \gls{hoa} synthesis, the spherical harmonic components of the sound field are decoded to a set of virtual loudspeakers positioned around the listener's head \cite{noisternig2003ambisonic, zaunschirm2018brir, otani2020binaural}. The decoded signals are then convolved with the corresponding \gls{hrtf} and summed to obtain the binaural signal that is presented to the listener through headphones. When the original sound field was analyzed using first-order \gls{shd}, directional sharpening can be achieved by using the \gls{sdm} to re-encode the sound field into higher-order \gls{shd} components, and subsequently decoding the higher-order \gls{shd} components to a set of virtual loudspeakers positioned around the listener's head \cite{frank2016spatial, zotter2019ambisonics}.

\subsection{Synthesis of historical worship space acoustics}

\noindent The reproduction of \gls{hws} acoustics using \gls{sfs} methods provides a unique and immersive experience for listeners and performers of historical music. Binaural synthesis of \gls{hws} acoustics has been used to recreate the acoustics of ancient Greek temples \cite{koutsivitis2005reproduction}, cathedrals \cite{postma2016virtual, suarez2016archaeoacoustics, alvarezmorales2017virtual, eley2021virtual, mullins2023immersive}, churches \cite{autio2021historically, lopezmochales2022experimental, berger2023exploring, alambeigi2024auralising}, ancient Egyptian temples \cite{warusfel2021assessing}, and mosques \cite{suarez2018virtual, berger2023exploring}.

Binaural synthesis renders a historical sound field for individual listeners. This has the advantage of providing a more controlled listening experience, as the listener is acoustically decoupled from the listening environment. However, for interactive \gls{sfs} scenarios, such as live performances, binaural synthesis may not be suitable as it does not provide a shared listening experience for multiple listeners, and may lead to a feeling of disconnection amongst musicians. In contrast, loudspeaker synthesis can be used to recreate the historical sound field over an extensive listening area, allowing for multiple listeners to experience the sound field as they move around the room. Loudspeaker synthesis has been used to recreate the acoustics of churches \cite{martellotta2008subjective}, cathedrals \cite{ canfielddafilou2019method, eley2021virtual}, and the Stonehenge monument \cite{fazenda2013recreating}.
 \section{Real-time Auralization}
\label{sec:review2024:real_time_auralization}

\noindent Real-time auralization of acoustics is an essential tool in the field of audio engineering, music production, and virtual and augmented reality applications. The goal of real-time auralization is to reproduce the acoustics of a given environment in real time, allowing the user to experience the sound of the environment as if they were physically present in that space.

In real-time auralization, the processing time of the audio signal must be kept within a desired latency budget to ensure that the synthesized sound field is perceived without noticeable delay. The latency budget can be defined as the maximum allowable delay between the input audio signal and the output sound field, and for block-based processing with a block size of $n_x$ samples, is defined as $n_x / f_s$, where $f_s$ denotes the sampling frequency of the synthesis system. For example, in a real-time auralization system with a block size of $n_x = 256$ samples and $f_s = 48$ $\si{\kilo\hertz}$, the maximum allowed processing latency would be $5.33$ $\si{\milli\second}$.

Environments with rich and complex acoustics, such as concert halls and \gls{hws}, are characterized by their long reverberation time. The acoustic synthesis of such spaces requires filters containing many filter taps, which can be computationally expensive to convolve with an audio signal under real-time conditions. To address this issue, various methods have been proposed to accelerate the convolution process, including uniform and non-uniform partitioned convolution, GPU acceleration, and \gls{rir} compression techniques. In addition to convolution-based methods, feedback delay networks and scattering delay networks have been proposed as alternative approaches to interactive auralization.

Finally, acoustic feedback cancellation is an important consideration in real-time auralization systems, as it can help to prevent unwanted feedback loops that can occur when the output of a loudspeaker is picked up by a microphone and re-amplified, causing unwanted artifacts in the reproduced sound such as howling or ringing \cite{vanwaterschoot2011fifty}. Acoustic feedback cancellation algorithms are designed to identify and suppress feedback signals in real time, allowing for stable and artifact-free auralization.

\subsection{Methods}
\subsubsection{Fast Convolution}

\noindent Convolution-based methods enable real-time auralization of room acoustics by convolving a dry audio signal, such as an instrument recording or a musical performance, with the filters derived from the synthesis methods outlined in the previous section. The resulting audio signal is then played back through a loudspeaker system or headphones, allowing the listener to experience the sound of the synthesized environment. Convolution-based methods are widely used in real-time auralization systems due to their accuracy and flexibility in modeling complex acoustic environments. However, the computational cost of convolving synthesis filters with audio signals in real time can lead to a prohibitively long processing delay, especially when synthesizing sound fields with long reverberation times in multichannel loudspeaker arrays. 

To illustrate the computational cost of convolution-based methods in the time-domain, consider the following example: for a room with a reverberation time of $10$ seconds, a 10-second-long \gls{rir} that is sampled at 48 \si{\kilo\hertz} requires $480,000$ filter taps to accurately model the sound field. When synthesizing the sound field in a loudspeaker array containing $24$ channels, the time-domain convolution of the audio signal with the synthesis filters requires approximately $11.5 \times 10^6$ multiply-accumulate operations per input sample. At a sampling rate of 48 \si{\kilo\hertz}, this corresponds to a computational load of $1.11$ TFLOPS, which is beyond the capabilities of most consumer-grade CPUs.

\paragraph{Frequency-Domain Convolution}

\noindent The computational cost of time-domain convolution can be reduced by performing the convolution in the frequency domain using the block-based \gls{fft} algorithm \cite{cooley1965algorithm}. In frequency-domain convolution, the audio signal is buffered in a block of $n_x$ samples, after which the input block and the synthesis filters are transformed into the frequency domain using an \gls{fft} size of $n_f$ frequency bins. When $n_x < n_f$, the input block is zero-padded to match the \gls{fft} size. The frequency-domain transformed input block and synthesis filters are then multiplied in the frequency domain, and the resulting product is transformed back to the time domain using the inverse \gls{fft} to obtain the output signal. A block-based \gls{ola} or \gls{ols} method can be used to obtain the final output block of length $n_x$ samples. The latency of frequency-domain convolution is determined by the block size $n_x$, and the computational cost is $O(n_f \log n_f)$ \cite{cooley1965algorithm}. The computational advantage of frequency-domain convolution over time-domain convolution increases with the block size $n_x$, however, this block size $n_x$ is upper bounded by the latency budget. Therefore, by choosing an appropriate \gls{fft} size, the computational cost of frequency-domain convolution can be lower than time-domain convolution. Nevertheless, the computational cost of frequency-domain convolution can still exceed the latency budget when processing large filter lengths, as the \gls{fft} size is lower bounded by the sum of the longest filter length and the analysis window length, minus one.

\paragraph{Partitioned Convolution}

\noindent To address the computational complexity of frequency-domain convolution with long synthesis filters, the filters can be decomposed into a set of shorter sub-filters that allow for frequency-domain convolution with an input signal within the latency budget. This approach, known as partitioned convolution, was first presented by Stockham in 1966 for processing large filters \cite{stockham1966high} and has been introduced for real-time \gls{fir} filtering in audio applications by Torger and Farina in 2001 \cite{torger2001realtime}. Two main approaches to partitioned convolution have been proposed: uniform partitioned convolution and non-uniform partitioned convolution.

In uniform partitioned convolution, a filter of length $n_h$ is decomposed into a set of $K$ sub-filters of length $n_k$, such that $n_h \leq K \cdot n_k$. A uniformly-partitioned input signal with block-length $n_x$ is then convolved with each sub-filter in the frequency domain using the \gls{fft} algorithm. Each sub-filter is offset by $k \cdot n_k$, where $k$ is the index of the sub-filter. The resulting partial convolutions are then summed using the \gls{ols} or \gls{ola} method to obtain the final output signal, which is outputted in blocks of length $n_x$ \cite{wefers2014partitioned}. When $n_x = n_k$, the delays of all sub-filters can be directly realized in the frequency domain \cite{wefers2014partitioned}.

In non-uniform partitioned convolution, the filter is decomposed into a set of $K$ sub-filters of varying lengths, allowing for more efficient use of the available computational resources. Since frequency-domain convolution becomes computationally more efficient as the \gls{fft} size increases, non-uniform partitioned convolution can be used to minimize the computational cost of the convolution process by choosing the optimal length of each sub-filter in the frequency domain. Since the latency-budget of each sub-filter is determined by $k \cdot n_x / f_s$, the length of each sub-filter can be adjusted to balance the computational cost of frequency-domain convolution with the latency budget. For later sub-filters, the length can be increased to reduce the computational cost of the convolution process, while for earlier sub-filters, the length can be reduced to ensure that the convolution process remains within the latency budget \cite{wefers2014partitioned}.

In partitioned convolution methods, the transformation from the time-domain to the frequency domain is performed only once for all sub-filters and input blocks, reducing the computational cost of the convolution process. For a comprehensive overview of uniform and non-uniform partitioning methods and their application to real-time auralization systems, the reader is referred to the work of Wefers \cite{wefers2014partitioned}.

Partitioned convolution methods decompose the filter into smaller sub-filters that can be concurrently processed with an input signal. This approach makes uniform and non-uniform partitioned convolution well-suited for parallel processing on \glspl{gpu}, which are optimized for processing large amounts of data in parallel. By delegating the convolution process to a GPU, the processing time can be significantly reduced \cite{wefers2010high}, enabling real-time auralization with long reverberation times in multichannel loudspeaker arrays.

\paragraph{Computational reduction techniques}
\noindent Computational reduction techniques aim to reduce the number of filter taps required to accurately model the sound field while maintaining the perceptual quality of the synthesized sound. Common techniques are thresholding, truncation, and compression of the synthesis filters. In thresholding, the filter taps with magnitudes below a certain threshold $\epsilon$ are set to zero, reducing the number of non-zero filter taps in the filter, and therefore reducing the required amount of addition and multiplication operations in the time-domain convolution process. In truncation, the filters are truncated, so that the convolution with an input signal can be performed within the desired latency budget \cite{wendt2014computationally, cadavid2022performance}. Compression of the synthesis filters can be achieved by exploiting the underlying structure of the filters. For example, it was shown in \cite{jalmby2021low} that filters modeling room acoustics, i.e., \glspl{rir}, exhibit a low-rank structure when rearranged into a matrix or tensor. This property can be used to compress the filters using low-rank matrix factorization techniques, reducing the number of filter taps \cite{jalmby2021low}. Moreover, the compressed filter representation can be directly used for fast convolution in the time domain \cite{jalmby2023fast, jalmby2024compression}. In \cite{jalmby2024multi}, this method was extended to multichannel \glspl{rir} measured in the same room, showing that the low-rank structure is shared across channels, which can be exploited to further reduce the computational cost of the convolution process. The authors applied this method to compress and achieve multichannel fast convolution of synthesis filters in real-time auralization systems \cite{rosseel2024low}. While low-rank compressed filters offer computational advantages, it is important to note that they can introduce artifacts in the synthesized sound field at high compression rates, specifically in the high-frequency range \cite{rosseel2024low}.

\paragraph{Filtered Velvet Noise}
\label{sec:review2024:real_time_velvet_noise}
\noindent Computational reduction can also be achieved by separating the late reverberation from the direct and early reflections of the sound field and synthesizing them separately. A computationally efficient method for modeling the late reverberation can be achieved using \gls{fvn} \cite{karjalainen2007reverberation, valimaki2017late}. Velvet noise is a type of sparse pseudo-random ternary sequence that contains only the values $-1$, $0$, and $1$. This inherent sparsity allows for efficient convolution, as the convolution process can be implemented using a series of additions and subtractions, rather than multiplications. To model the late reverberation of a sound field using \gls{fvn}, the late reverberation is first divided into segments, each approximated by an independent \gls{fvn} sequence. To obtain the synthesized reverberation, an input signal is filtered with the cascading \gls{fvn} sequences, and a coloration filter and an associated gain factor are applied to approximate the frequency-dependent decay of the reverberation in the time domain. The resulting \gls{fvn}-filtered sequences are summed to obtain the final output \cite{karjalainen2007reverberation, valimaki2017late}. Recent advancements have extended \gls{fvn} to incorporate parameter estimation using machine learning, enabling automatic design of \gls{fvn} sequences to match the energy decay profile of a given acoustic environment \cite{lee2022differentiable}. While \gls{fvn} efficiently reproduces an approximation of the late reverberation of a sound field, noticeable differences can still exist between measured and synthesized reverberation \cite{valimaki2017late}.

\subsubsection{Delay Networks}

\noindent Due to the computational complexity of convolution-based methods, alternative approaches to achieve real-time auralization have been proposed. Delay networks are types of recursive filters that can be used to efficiently synthesize artificial reverberation with low computational cost \cite{stautner1982designing, jot1991digital, valimaki2012fifty}. An \gls{fdn} is a recursive filter that consists of a network of delay lines, a set of input and output gains, and a unitary feedback matrix that models the feedback path of the filter. The feedback matrix determines the frequency response of the \gls{fdn} and can be designed to model the frequency-dependent sound energy decay of a given acoustic environment \cite{valimaki2012fifty, bai2015late, shen2020data, lyster2022differentiable, bona2022automatic, ibnyahya2022method}.

The output of the \gls{fdn} is obtained by summing the delayed and attenuated signals from the delay lines, which creates a dense and diffuse reverberation effect. \glspl{fdn} are widely used in real-time auralization systems due to their low computational cost and ability to model long reverberation times \cite{valimaki2012fifty}.

The spectrum of artificial reverberation synthesized by an \gls{fdn} is often colored due to resonating modes. To address this issue, differentiable \glspl{fdn} have been proposed, which optimize the feedback matrix, as well as the input and output gains, to achieve a near-colorless artificial reverberation \cite{dalsanto2023differentiable}.

While \glspl{fdn} offer a computationally efficient method that is simple to design, which can achieve high quality artificial reverberation, they lack the ability to accurately reproduce the direct and early reflections of a space. Therefore, hybrid methods that combine convolution-based methods with \glspl{fdn} have been proposed. In \cite{carpentier2014hybrid}, a hybrid reverberation processor was introduced that reproduces the early reflections of a space using convolution-based methods and synthesizes the late reverberation using an \gls{fdn} that is automatically adjusted to match the energy decay profile of the space \cite{carpentier2014hybrid}.

Another method to synthesize the early reflections and room modes using a recursive filter is the \gls{sdn}, which was proposed by De Sena \textit{et al.} \cite{desena2015efficient}. \glspl{sdn} are artificial reverberators that model the early reflections and room modes of a space using a network of scattering nodes. To model the acoustic path between a source and receiver in a room using \glspl{sdn}, each reflective surface in the room is represented by a single scattering node that is placed at the location on the surface corresponding to the first-order reflection point. The scattering nodes are connected with the source, and with each other using bidirectional delay lines that model the sound propagation delay between the nodes and the absorption of the sound energy by the reflective surfaces \cite{desena2015efficient}. The receiver node is connected to each scattering node using a unidirectional delay line that models the sound propagation from each reflective surface to the receiver. The scattering matrix determines how the energy is circulated within the network \cite{desena2015efficient}. By positioning the scatter nodes at the locations of the first-order reflections, higher-order reflections and room modes are implicitly modeled by the scattering network, allowing for the synthesis of a realistic sound field with low computational cost \cite{desena2015efficient}.

\subsubsection{Acoustic Feedback Cancellation}
\label{sec:review2024:real_time_feedback_cancellation}

\noindent Real-time interactive auralization systems use microphones to capture sound sources in the environment. These microphone signals are typically processed by a conditioner function $\mathbf{G}$, convolved with the synthesis filters $\mathbf{H}$, and played back through loudspeakers to create the synthesized sound field. Acoustic feedback occurs when the output of the auralization process is played back through a loudspeaker system and is again picked up by the microphones, creating a feedback loop that can lead to unwanted artifacts. Acoustic feedback cancellation algorithms are designed to identify and suppress feedback signals, allowing for stable and artifact-free auralization \cite{vanwaterschoot2011fifty}.

\begin{figure}[ht]
    \centering
    \includegraphics[width=0.8\columnwidth]{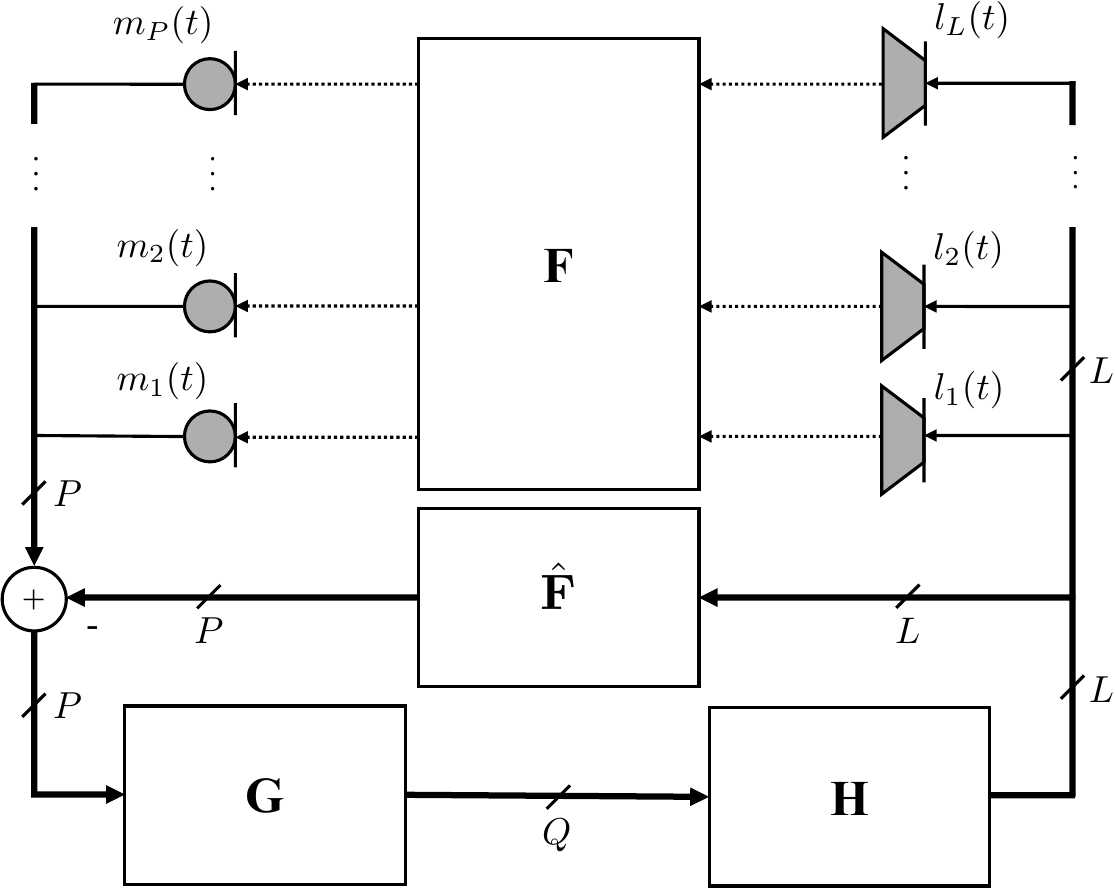}
    \caption{Block diagram of a real-time auralization system containing $L$ loudspeakers and $P$ microphones. The microphone signals are first preprocessed by a conditioner function $\mathbf{G}$, after which they are convolved with the synthesis filters $\mathbf{H}$ to create the synthesized sound field. Acoustic feedback cancellation is implemented using the multichannel feedback path estimate $\mathbf{\hat{F}}$ and subtracting the resulting feedback signals from the microphone signals, to compensate for the acoustic feedback path $\mathbf{F}$.}
    \label{fig:review2024:aur_system_w_feedback}
\end{figure}

Acoustic feedback cancellation is a well-explored area within public address systems, characterized by unknown and variable feedback paths $\mathbf{F}$ due to the movement of sound sources and microphones \cite{vanwaterschoot2011fifty}. In this area, several techniques have been proposed for mitigating acoustic feedback, including phase modulation \cite{schroeder1964improvement, nielsen1999performance, poleti2004stability}, and notch filters \cite{er1994dsp, gilcacho2009regularized}. However, these methods may introduce sound artifacts, including coloration and distortion, which degrade the perceptual quality of the auralization. Alternatively, adaptive filtering techniques can be employed to estimate the multichannel feedback paths $\mathbf{F}$. The estimated feedback signal is then subtracted from the microphone signals to obtain the feedback-compensated signal \cite{vanwaterschoot2004performance, spriet2006adaptive, vanwaterschoot2009adaptive}. For an exhaustive overview of acoustic feedback cancellation methods in public address systems, the reader is referred to the work of van Waterschoot \textit{et al.} \cite{vanwaterschoot2011fifty}.

In the context of real-time auralization systems, where the loudspeaker and microphone configurations are typically fixed, the multichannel feedback paths $\mathbf{F}$ do not rapidly change over time. This allows for the estimation of the feedback paths $\mathbf{F}$ using system identification techniques in a calibration phase and use the estimated feedback paths to cancel the feedback signals in real time during auralization without altering the perceptual quality of the synthesized sound field. In Fig. \ref{fig:review2024:aur_system_w_feedback}, the block diagram of a real-time auralization system with acoustic feedback cancellation is shown. The microphone signals are first preprocessed by a conditioner function $\mathbf{G}$, after which they are convolved with the synthesis filters $\mathbf{H}$ to create the synthesized sound field. Acoustic feedback cancellation is implemented by estimating the multichannel feedback path $\mathbf{F}$ and subtracting the estimated feedback signals from the microphone signals.

In \cite{abel2018feedback}, an acoustic feedback cancellation algorithm was proposed for real-time auralization systems. The method designs a multichannel feedback path estimate $\mathbf{\hat{F}}$ that incorporates robustness to measurement errors, model uncertainties, and slight variations in the feedback paths due to moving performers or participants. For a single loudspeaker and microphone, the robust time-domain feedback path estimate $\hat{f}(t)$ is obtained by applying a time-frequency weighting function $w(t, f_b)$ to the measured feedback path $\tilde{f}(t, f_b)$, where $f_b$ denotes the frequency bin in the time-frequency representation, i.e.,

\begin{equation}
    \hat{f}(t) = \sum_{b=1}^{B} \tilde{f}(t, f_b) \cdot w(t, f_b),
\end{equation}

\noindent where $B$ represents the number of frequency bins in the time-frequency representation of the feedback path. The weighting function $w(t, f_b)$ is designed to suppress the feedback path estimate in regions where the feedback path is not well known by using a Wiener-like weighting, i.e.,

\begin{equation}
    w(t, f_b) = \frac{f^2(t, f_b)}{f^2(t, f_b) + \sigma^2(t, f_b)},
\end{equation}

\noindent where $f(t, f_b)$ denotes the true feedback path and $\sigma^2(t, f_b)$ denotes the variance of the feedback path estimate \cite{abel2018feedback}. The true feedback path $f(t, f_b)$ and variance $\sigma^2(t, f_b)$ can be estimated by repeated measurements of the feedback path in the calibration phase. The true feedback path $f(t, f_b)$ is estimated by averaging the measured feedback paths over time, while the variance $\sigma^2(t, f_b)$ is estimated by calculating the variance of the measured feedback paths over time. Additionally, $f^2(t, f_b)$ is smoothed over time and frequency to obtain a smooth windowing function \cite{abel2018feedback}.

The robust feedback path estimate $\hat{f}(t, f_b)$ is then used to design a feedback cancellation filter that is applied to the microphone signals to obtain the feedback-compensated signal. The method was shown to achieve robust feedback suppression greater than $20$ dB in real-time multichannel auralization systems \cite{abel2018feedback}. The method has been used to design auralization systems in a variety of applications, including recording studios \cite{baran2023optimization, grazioli2023preliminary}, \gls{vr} for historical preservation \cite{berger2023exploring}, and a series of art installations \cite{callery2023convolution}.

\subsection{Real-time auralization of historical worship space acoustics}

\noindent Real-time auralization of \gls{hws} is a challenging task due to the rich acoustics and long reverberation times by which these spaces are characterized. The acoustics of \gls{hws} are of particular interest to musicologists, musicians, and audio engineers, as they provide a unique sonic environment in which music and liturgy were composed and performed. Real-time auralization of \gls{hws} allows for the interactive exploration of the acoustics of these spaces, enabling musicians to experience the sound of the space as if they were physically present in it, and to investigate the intrinsic relationship between musical performance and the space in which it is performed \cite{schiltz2003church, ueno2003experimental, bassuet2004acoustics, ueno2010effect}.

The real-time auralization of \gls{hws} using convolution-based methods has been proposed in several studies. Uniformly partitioned convolution methods have been used to synthesize the acoustics of the Cath{\'e}drale Notre-Dame de Paris \cite{mullins2023immersive} and the acoustics of ancient Greek spaces \cite{koutsivitis2005reproduction}.

A non-uniform partitioned convolution scheme was used by Eley \textit{et al.} to synthesize the early reflections of Cath{\'e}drale Notre-Dame de Paris in real time \cite{eley2021virtual}. The late and diffuse reverberation of the cathedral was synthesized using an \gls{fdn} model. The authors showed that the combination of non-uniform partitioned convolution and \gls{fdn} synthesis can achieve a realistic and immersive auralization of the cathedral in real time \cite{eley2021virtual}.

In real-time binaural synthesis of \gls{hws}, head tracking techniques have been proposed to enhance the spatial realism of the auralization. Head tracking allows the listener to move their head in the virtual space, changing the perceived direction of the sound sources and reflections. Postma \textit{et al.} \cite{postma2016virtual} used head tracking in a \gls{vr} system to enhance the spatial realism of the auralization of the Cath{\'e}drale Notre-Dame de Paris. Head tracking was also used by Berger \textit{et al.} \cite{berger2023exploring} in a \gls{vr} system which enabled the listener to experience the acoustics of ancient caves, mosques, and churches in 9-\gls{dof} \gls{vr}.

In \cite{canfielddafilou2019method}, real-time loudspeaker-based auralization of the Chiesa di Sant'Aniceto, a historical church in Rome, was achieved by implementing the feedback-canceling reverberator. 

Real-time auralization of \gls{hws} can also be implemented in the Max/MSP software \cite{maxmsp2024}. Max/MSP software is widely used for real-time audio applications, and has been used for the real-time binaural auralization of an ancient Egyptian temple \cite{warusfel2021assessing}, and the Notre-Dame de Paris \cite{mullins2023immersive}.

A summary of software enabling the real-time synthesis of room acoustics and their use in reproducing the acoustics of \gls{hws} is provided in Table \ref{tab:review2024:real_time_auralization_software}.

\begin{table}[h!]
    \centering
    \caption{Software that enables real-time synthesis of room acoustics and their use in reproducing the acoustics of \gls{hws}.}
    \begin{NiceTabular}{X[l, 3] X[l, 3] X[l, 3]}
        \toprule
        Software & License & Use in \gls{hws} \\
        \midrule
        Max/MSP \cite{maxmsp2024} & Commercial & \cite{postma2016virtual, eley2021virtual, warusfel2021assessing, mullins2023immersive} \\
        Pure Data \cite{puckette1997pure} & BSD-3-Clause & \\
        \bottomrule
    \end{NiceTabular}
    \label{tab:review2024:real_time_auralization_software}
\end{table} \section{Perceptual Evaluation}
\label{sec:review2024:perceptual_evaluation}
\noindent Loudspeaker and headphone-based sound field synthesis methods aim to accurately synthesize an acoustic environment in a reproduction room. The original sound field in the room is however, immensely complex, making the true physical reproduction of the sound field using a finite number of microphones and loudspeakers challenging. Luckily, the human auditory system is prone to perceptual illusions, which can be exploited to create a plausible reconstruction of the virtual sound field in the reproduction room. The perceptual evaluation of sound field synthesis methods is crucial to assess the perceived quality of the synthesized sound field. In this section, we will discuss the various perceptual evaluation methods in the context of sound field auralization systems.

\subsection{Methods}

The reproduction quality of auralization systems can be evaluated using objective psychoacoustic-based methods, subjective methods, or sensory evaluation methods.

\subsubsection{Objective psychoacoustic-based evaluation methods}
\label{sec:review2024:perceptual_eval_methods_psych}

 Objective psychoacoustic-based evaluation methods are based on the human auditory system's perception of sound. These methods involve the use of psychoacoustic models that predict the perceived quality of the synthesized sound field based on the physical properties of the sound field \cite{beradek2004concert}. These physical properties can be measured from acoustic measurements. The most common objective evaluation parameters used in the context of sound field synthesis are \gls{rt}, \gls{edt}, Clarity, Definition, Sound Strength, \gls{asw}, Listener Envelopment, \gls{icc}, Early Support, and \gls{sti}, all of which are defined in the ISO3382-1:2009 standard \cite{ISO3382-1}.

 The perceptual evaluation of acoustics for a specific application can be performed by aggregating 
 the objective evaluation parameters into a single index that represents the perceived quality of 
 the sound field for specific applications. This has been done to perceptually evaluate concert 
 hall acoustics \cite{ando1998architectural, beradek2004concert} by applying a weighting factor to each objective evaluation parameter based on the perceived importance of the parameter for the application.

\subsubsection{Subjective evaluation methods}

Subjective evaluation methods involve human listeners who evaluate the synthesized sound field based on their perception by listening or interacting with the system and providing feedback through questionnaires \cite{barron1988subjective, patynen2014amplitude, tervo2014preferences, huttenmeister2016perceived, ahrens2019perceptual} or interviews \cite{ueno2003experimental,ueno2010effect, amengual2017investigations, tomasetti2023playing, demuynke2024ears}. The resulting data from the subjective evaluation methods can be processed using statistical tools to analyze the perceived quality of the synthesized sound field \cite{mendon2018statistical}.

Perceptual evaluation methods can be broadly divided into two categories: \textit{Comparative} and \textit{Descriptive} methods \cite{bech2006perceptual}. Comparative methods evaluate aspects of stimuli in comparison to a reference stimulus \cite{amengual2016physical, coleman2024exploring}. For example, in preference tests, the listener is asked to compare two stimuli and indicate a preference for one stimulus \cite{schroeder1974comparative, ando1979subjective, ueno2003experimental, patynen2014amplitude, tervo2014preferences, ando2014concert,lokki2016concert, ahrens2019perceptual}. Descriptive methods involve evaluating each stimulus by ranking a set of descriptive attributes based on the perceived intensity of each attribute for every stimulus. The type of evaluation method that is used depends on the research question and the context of the study \cite{bech2006perceptual}. For an overview of perceptual evaluation methods used in the context of audio evaluation, the reader is referred to \cite{bech2006perceptual}.

\subsubsection{Sensory Evaluation Methods}

While comparative and descriptive evaluation methods are commonly used in the subjective evaluation of audio, sensory evaluation methods offer a more structured approach to evaluate sensory data. These methods were first introduced in the food industry to evaluate the sensory properties of food and beverages \cite{williams1985comparison, lawless1999sensory}, and have since been used in the perceptual evaluation of reproduced concert hall acoustics \cite{lokki2010auditorium, lokki2011concert, lokki2012disentangling}, car cabin acoustics \cite{kaplanis2017perceptual, kaplanis2017rapid, camilleri2019evaluation}, and small room acoustics \cite{kaplanis2014perception, kaplanis2019perception, lachenmayr2023chamber}. 

In sensory evaluation methods, a set of descriptors, also called attributes, should be carefully selected to accurately describe the perceived sensation across a set of stimuli. When experienced or expert assessors are involved in the evaluation, the attributes can be elicited using \gls{cvp} where the group of assessors agrees on a common set of attributes that best describe the perceived sensational differences between all stimuli before the evaluation \cite{lindau2014spatial}. However, it can be challenging and time-consuming to elicit a common set of attributes that accurately represent the perceived sensation of the stimuli, especially for large groups of assessors. As an alternative, \gls{ivp} can be used, where the assessors describe the perceived sensations using their own vocabulary \cite{lorho2005individual}. In \gls{ivp}, the usage of each elicited attribute may not be consistent across assessors, which can make the interpretation of the results challenging. To address this issue, statistical methods, such as \gls{pca} and \gls{mfa}, can be used to extract the meaning of the attributes and to perform a correlation analysis between the attributes and the stimuli \cite{lorho2005individual}. Moreover, the relationships between preference ratings, sensory evaluation methods, and acoustic measurements can be visualized using \gls{mfa} \cite{kuusinen2014relationships}, providing valuable insights into the perceived quality of the stimuli. The attribute elicitation phase is followed by the evaluation phase, where the assessors evaluate the stimuli based on the elicited attributes. After the experiment, the data collected from the assessors is analyzed using statistical methods, such as \gls{gpa}, \gls{mfa}, \gls{pca}, and ANCOVA, to evaluate the perceived quality of the stimuli \cite{bech2006perceptual}.

A widely recognized method for evaluating the sensory properties of food and beverages is \gls{qda}, where a panel of trained assessors evaluates the sensory properties of the stimuli using \gls{cvp} \cite{stone2004sensory}. The assessors undergo training to recognize and describe the sensory properties of the stimuli under evaluation, and the data collected from the assessors are analyzed using statistical methods to evaluate the perceived quality of the stimuli. However, \gls{qda} is time-consuming and requires a panel of trained assessors, which can be expensive and difficult to maintain. This limitation has restricted its application in the perceptual evaluation of audio reproduction systems. To address the limitations of \gls{qda}, faster sensory evaluation methods that do not require extensive training of the participants such as \gls{fcp} and \gls{fp} have been developed. These methods, which rely on \gls{ivp}, enable the quick evaluation of sensory properties using untrained or semi-trained assessors \cite{lawless1999sensory, dairou2002comparison}. 

During the elicitation phase of \gls{fcp} and \gls{fp}, each assessor is asked to identify the descriptors that best describe the perceived sensation of the presented stimuli. These descriptors are individually elicited without the need for a common definition or agreement among the assessors. In \gls{fcp}, assessors then rate the set of attributes based on the perceived intensity of each attribute for every stimulus. After the experiment, the data is analyzed using \gls{gpa} to align the data from each individual assessor into a common space, resulting in a consensus profile that represents the perceived quality of the stimuli. In \gls{fp}, instead of rating the attributes, the assessors are asked to rank the set of attributes based on the perceived intensity of each attribute for every stimulus, and the data collected from the assessors is analyzed using \gls{mfa}. The advantage of \gls{fcp} and \gls{fp} is that they are quick and easy to perform without requiring a panel of trained assessors. Moreover, while the results obtained from \gls{fcp} and \gls{fp} may not be as accurate or complete as those obtained from \gls{qda}, they provide similar insights into the perceived quality of the stimuli \cite{lawless1999sensory, dairou2002comparison,mendon2018statistical}. According to one comparative study, the \gls{fp} method was found to display a stronger discrimination capability than \gls{qda} and \gls{fcp} in taste evaluations \cite{liu2018comparison}.

Given the efficiency and ease of use of \gls{ivp} methods, they have been used in the perceptual evaluation of auditorium and concert hall acoustics \cite{lokki2010auditorium, lokki2011concert, lokki2012disentangling}, car cabin acoustics \cite{kaplanis2017rapid,kaplanis2017perceptual}, small room acoustics \cite{kaplanis2019perception}.

The auditory sensory evaluation methods described above are non-interactive, meaning the assessors evaluating the stimuli did not interact with the stimuli during the evaluation. Interactive sensory evaluation, on the other hand, involve the assessors interacting with the auditory stimuli to evaluate the perceived quality of the stimuli. This interaction provides a more realistic evaluation of the stimuli, as the assessors can evaluate the stimuli based on their perception of the stimuli in a real-world scenario. In the context of interactive sensory evaluation for audio reproduction systems, \gls{fp} has been used to evaluate the perceived quality of virtual acoustic environments \cite{rosseel2023interactive}. In \cite{amengual2019analysis}, Dual Multiple \gls{mfa} was used to evaluate the perceived quality of trumpet performance adjustments in different virtual acoustic environments. Tomasetti \textit{et al.} \cite{tomasetti2023playing} compared musicians' preferences for collaborative music performance using binaural synthesis with head tracking over a conventional stereo setup. Similar research was conducted in \cite{chen2024enhancing}, where musicians' preference for binaural audio with head tracking over stereo was evaluated for digital piano performance.

In \cite{fargeot2023perceptual}, the perceptual localization performance of virtual sources was evaluated under real and reproduced acoustic conditions using a $4$th order 3D \gls{hoa} system. Participants were asked to localize virtual acoustic sources in a \gls{vr} environment by indicating the perceived \gls{doa}, apparent source width, and perceived distance of the virtual sources using a \gls{vr} pointer interface.

\subsection{Perceptual evaluation in historical worship spaces}

The acoustic quality of \gls{hws} can be evaluated using objective psychoacoustic-based evaluation parameters, as described in Section \ref{sec:review2024:perceptual_eval_methods_psych}. In \cite{engel2007index}, a synthetic index, an index that is based on a combination of multiple partial indices, was proposed to assess the acoustic properties of sacral buildings. The proposed index was a function of five partial indices, each representing a specific aspect of the acoustic quality of the building. Kosa{\l}a proposed a new formula for calculating the single number rating of the acoustic quality of church interiors \cite{kosala2011single} based on a combination of three partial indices using the \gls{svd} method. In \cite{berardi2012double}, a double synthetic index was proposed to evaluate the acoustics of churches. The double synthetic index separates the evaluation of the acoustics of churches into two metrics: an index for the perception of music, and an index for the perception of speech \cite{berardi2012double}.

An acoustic quality assessment was carried out by Kosa{\l}a \cite{kosala2016calculation} in the St. Elizabeth of Hungary church in Jaworzno, Poland.  The acoustic properties of the church were evaluated using two indices, namely the Global Acoustic Properties Index (GAP) \cite{kosala2013assessing} and the Global Index (GI) \cite{kosala2014comparative}. These acoustic evaluation methods were also used in the assessment of Orthodox churches in Poland \cite{kosala2018index}.

A widely used perceptual evaluation method for \gls{hws} is the measurement of objective acoustic parameters. Common parameters include \gls{rt}, \gls{edt}, clarity, definition, sound strength, sound level pressure, \gls{sti}, \gls{asw}, and \gls{icc}. These parameters offer a simple and effective way to evaluate the acoustics of \gls{hws}, as they can be directly calculated from \gls{rir} measurements. Objective acoustic parameters have been used to perceptually evaluate the acoustics of various \gls{hws}, including churches \cite{desarnaulds2002church, berardi2013simulation, elicio2015acoustics, brezina2015measurement, postma2015calibrated, postma2015creation, alonso2016room, zhao2016preliminary, alvarezmorales2017virtual, sender2018virtual, tronchin2020evaluation, alberdi2021acoustic, alberi2021evolutionary, sukaj2022byzantine}, cathedrals \cite{pedrero2014acoustical, alonso2014acoustic, suarez2015intangible, iannace2016acoustic, postma2016perceptive, martellotta2018investigation, alvarezmorales2020acoustic, canfielddafilou2024voices, demuynke2024ears}, mosques \cite{sügül2013impact, kavraz2016acoustic, suarez2018virtual, syamsiyah2018sound, kitapci2021acoustic, sukaj2021two}, and Buddhist temples \cite{soeta2013measurement, zhang2020effects}. 

Subjective evaluation methods have also been used to evaluate the perceived quality of the acoustics in \gls{hws}. In \cite{martellotta2008subjective}, a subjective evaluation of preferred listening conditions in nine Italian Catholic churches was performed. A total of 143 participants were included in the evaluation, where they were asked to state their preference for different binaural auralizations of church acoustics convolved with musical excerpts. The musical excerpts ranged from Gregorian chant to symphonic music. The results of the listening tests were analyzed by applying Factor Analysis to determine the key acoustic parameters that influenced the participants' preferences \cite{martellotta2008subjective}.

A subjective listening test conducted by Postma and Katz in 2016 \cite{postma2016perceptive} compared the perceptual similarity of measured and simulated binaural auralizations of the acoustics of the Saint-Germain-des-Pr{\'e}s abbey church, the Cath{\'e}drale Notre-Dame de Paris, and the Th{\'e}{\^a}tre de l'Ath{\'e}n{\'e}e, France. The participants were asked to compare the auralizations by rating the similarity of eight predefined perceptual attributes for each pair of auralizations \cite{postma2016perceptive}.

In \cite{zhang2016soundscape}, the perceptual evaluation of Han Chinese Buddhist temples was carried out. The evaluation involved a series of questionnaires which gauged the participants' comfort and preference for the soundscape at four temples. The soundscapes of each temple was measured and compared to the participants' responses to the questionnaires to identify the characteristics of the soundscapes that influenced the participants' comfort and preference \cite{zhang2016soundscape}.

The spatial impression and directional resolution of the reproduced acoustics of the St. Andrew's church, Lyddington, UK, was investigated by Frank and Zotter \cite{frank2016spatial}. A perceptual evaluation was done to compare the performance of acoustic reproduction with first-order spherical harmonics, and the \gls{sdm}-enhanced spherical harmonic representations of the sound field. Two listening tests were organized. The first listening test was used to determine the size of the perceptual sweet spot by asking participants to localize the reproduced sound at different locations in the reproduction setup. The second listening experiment used a \gls{mushra}-like graphical interface where the participants could rate the perceived spatial depth mapping for each auralization \cite{frank2016spatial}.

Subjective listening tests have also been carried out by {\'A}lvarez-Morales \textit{et al.} in 2017 \cite{alvarezmorales2017virtual} to evaluate the visual and acoustic experience of a reproduced \gls{vr} environment of the Cathedral of Seville, Spain. The evaluation consisted of three tests: a blind acoustic perception test, a visual virtualization test, and a combined visual and acoustic test. The participants were asked to rank the perceived quality of the visual and acoustic experience of the \gls{vr} environment \cite{alvarezmorales2017virtual}. For each test, the participants were asked to rank a set of predefined attributes based on the perceived quality of the \gls{vr} environment.

A perceptual assessment of reverberation in Spanish cathedrals was carried out by Gir{\'o}n \textit{et al.} in 2020 \cite{giron2020assessment}. A series of binaural listening tests with 42 participants were carried out to evaluate the perceived reverberation in the cathedrals. The participants had to compare the perceived reverberation between sets of two cathedrals. The results indicated that a reverberation time between four and seven seconds was indistinguishable for the listeners \cite{giron2020assessment}.

A recent study by L{\'o}pez-Mochales \textit{et al.} \cite{lopezmochales2022experimental} investigated the enhancement of feelings of transcendence, tenderness, and expressiveness by music in Christian liturgical spaces. The study involved a series of listening tests of music excerpts, one liturgical piece and one secular piece, that were presented to the listeners in either raw format, or an auralized format with different church acoustics. Using these excerpts, two listening tests were performed: one where the participants had to rate a set of attributes based on the perceived expressiveness and beauty of the music, and one preference test where the participants had to indicate their preference for the music excerpts. The results were analyzed using a factorial ANCOVA analysis to evaluate the effect of the church acoustics on the perceived expressiveness and beauty of the music \cite{lopezmochales2022experimental}.

The perceptual evaluation of a real-time auralization system reproducing the acoustics of the Chiesa di Sant' Aniceto in Rome, Italy, was carried out by Canfield-Dafilou \textit{et al.} in 2019 \cite{canfielddafilou2019method}. The evaluation involved two choirs, each choir consisting of three singers, that were asked to perform four verses of a Catholic hymn inside the auralization system. The auralization system reproduced the acoustics of the church in real-time, while varying the reverberation time during the performance. After the performance, the singers were asked to fill in an informal survey to evaluate the perceived quality of the auralization system \cite{canfielddafilou2019method}.

An interactive real-time auralization system for reproducing Cath{\'e}drale Notre-Dame de Paris' acoustics was validated using a perceptual analysis \cite{eley2021virtual}. The perceptual analysis was carried out by four singers, all members of a medieval ensemble with experience performing in the Cath{\'e}drale Notre-Dame de Paris. The singers were asked to validate the reproduction system by performing excerpts of medieval repertoire in the virtual reproduction space in both binaural and multichannel loudspeaker configurations. The singers' feedback was collected using questionnaires which evaluated the similarity of the virtual acoustics to the real acoustics of the cathedral \cite{eley2021virtual}. It should be noted that the participants could only base their comparison on their memory of the cathedral's acoustics \cite{eley2021virtual}, which might affect the reliability of the perceptual evaluation.

The relationship between ancient music practices and the acoustics of the space in which it is performed was investigated using an immersive binaural auralization system that was developed and presented by Mullins and Katz \cite{mullins2023immersive}. A subjective analysis in the form of an interactive choir performance was conducted to evaluate the suitability of various acoustical conditions for the performance of ancient music. The choir was asked to perform short excerpts of choral music using the interactive auralization system, and to rate the ease, suitability, and difficulty of performance after each excerpt \cite{mullins2023immersive}. \section{Example Case Study of the Nassau Chapel}
\label{sec:review2024:case_study}

\subsection{Introduction}

\noindent In this section, we present a case study of the preservation and auralization of the Nassau Chapel, also called St. George's Chapel, a 16th century Gothic style chapel located in Brussels, Belgium. The chapel was originally founded by Willem van Duvenvoorde in 1344 and was rebuilt in the early 16th century by the Nassau family as part of the Palace of Nassau \cite{vannieuwenhuyze2010brusselse}. The chapel was used as a public place of worship and services were held in the chapel throughout the Ancien R{\'e}gime. Today, the fully restored Nassau Chapel is incorporated into the buildings of the Royal Library of Belgium where it serves as an exhibition space displaying the manuscript collection of the Burgundian dukes \cite{kbr2024}.

The aim of the case study is twofold. First, we aim to digitally preserve the acoustics of the Nassau Chapel using room acoustic measurements and \gls{sfa} to capture the spatial and temporal characteristics of the sound field in the chapel. Second, we aim to use the measured data in a real-time auralization system that serves as an interactive rehearsal space for early music performers to explore the relationship between early music performances and the acoustics of the chapel.

\begin{figure}[t]
    \captionsetup[subfigure]{labelformat=empty}
    \begin{subfigure}[t]{0.49\columnwidth}
        \includegraphics[width=\columnwidth]{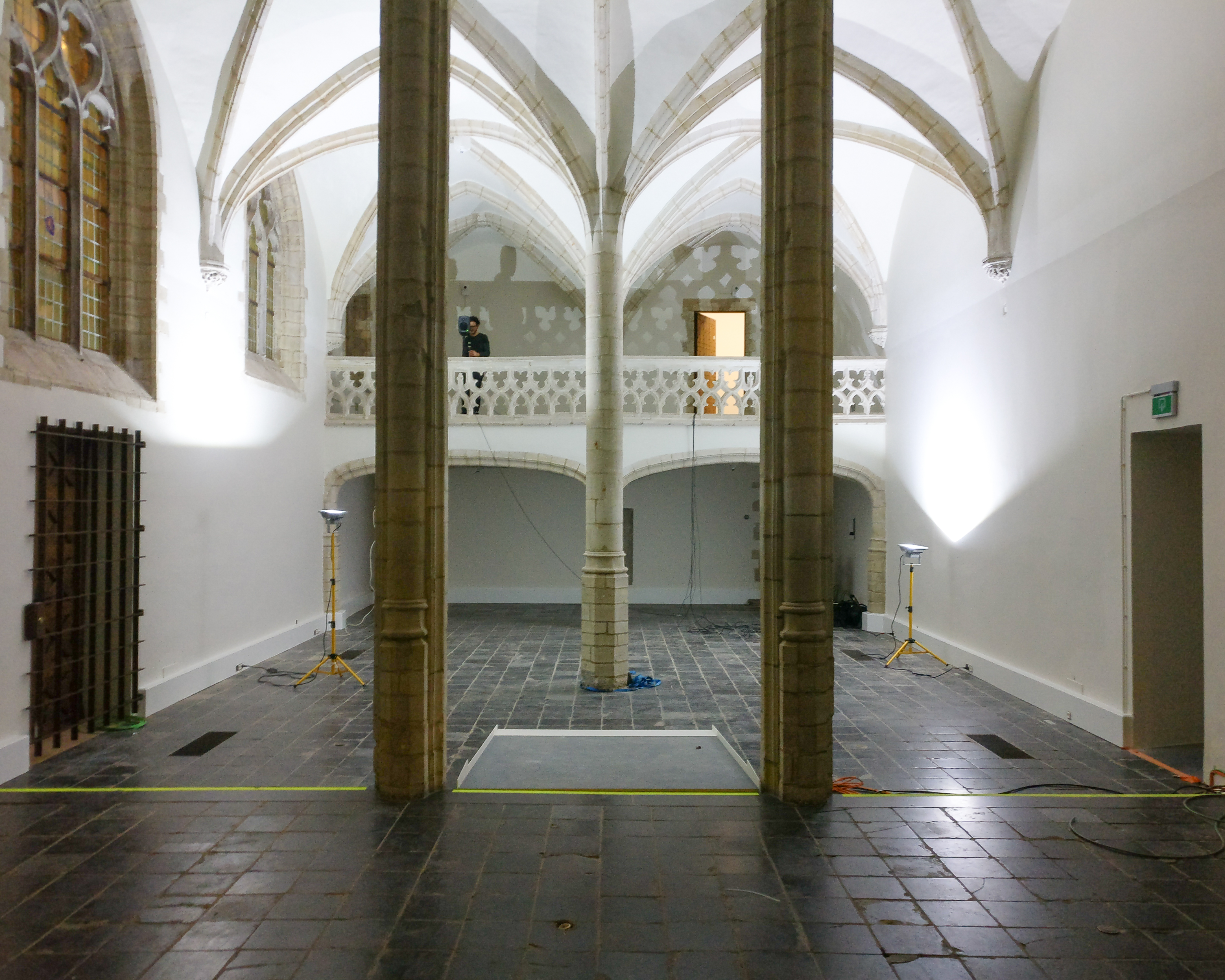}
        \caption{(a)}
    \end{subfigure}
    \hfill
    \begin{subfigure}[t]{0.49\columnwidth}
        \includegraphics[width=\columnwidth]{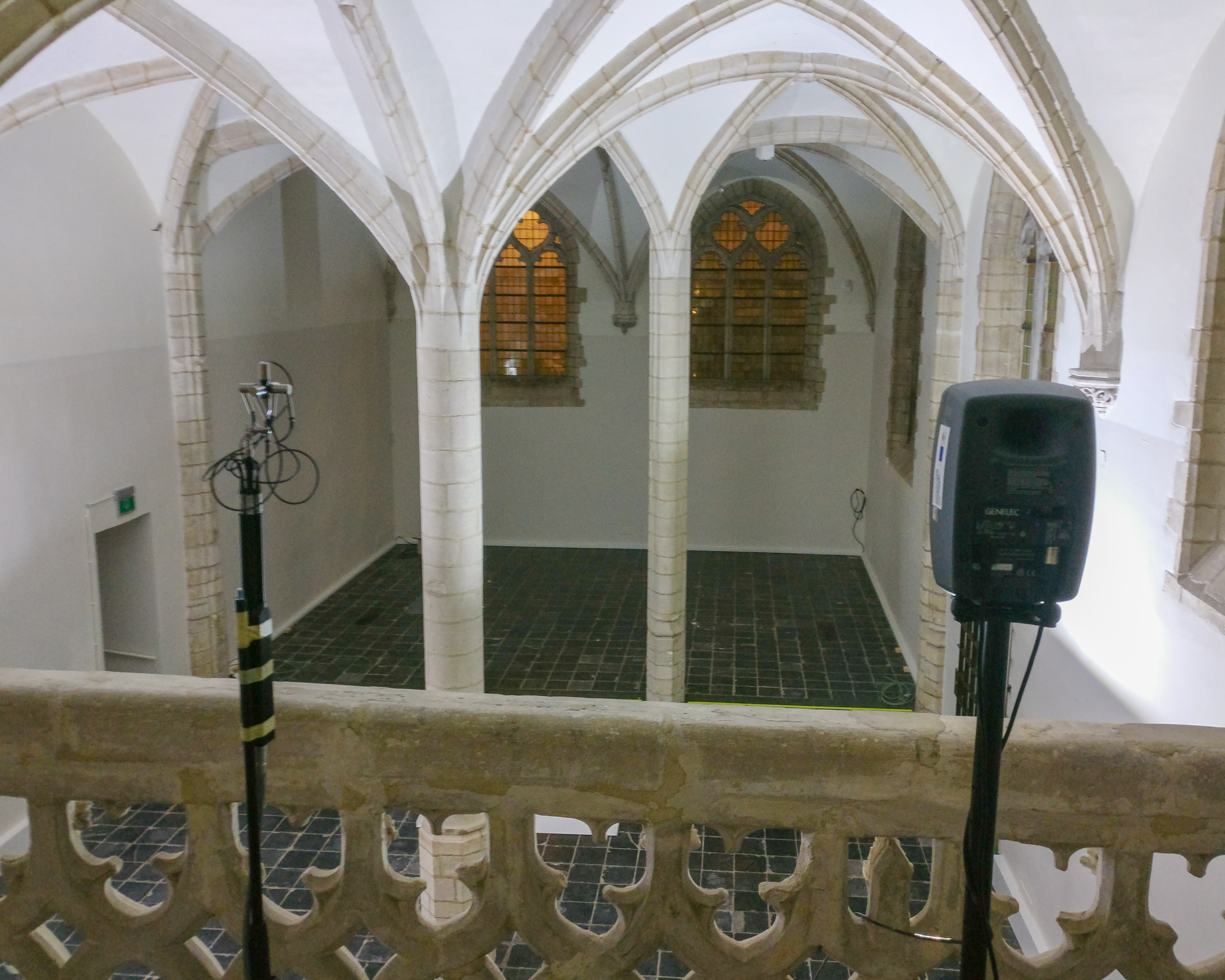}
        \caption{(b)}
    \end{subfigure}
    \caption{(a) View from the chapel floor looking towards the choir loft. (b) View from the choir loft looking onto the chapel floor. The loudspeaker used during the acoustic measurements is positioned on the right side of the choir loft.}
    \label{fig:review2024:nassau_chapel_view}
\end{figure}

The case study of the Nassau Chapel is divided into five main sections, according to the components of the auralization process presented in Fig. \ref{fig:review2024:overview}. Section \ref{sec:review2024:nassau_room_acoustic_acquisition} describes the room acoustic acquisition of the Nassau Chapel, including the measurement setup and procedure used to capture the RIRs at several positions in the chapel. Section \ref{sec:review2024:nassau_sound_field_analysis} presents a spatio-temporal analysis of the acoustic measurements made in the chapel using the \gls{sdm} as a tool to extract the spatial and temporal characteristics of the sound field. In Section \ref{sec:review2024:nassau_sound_field_synthesis}, the reproduction setup that is used to reproduce the sound field is presented. The Nassau Chapel's acoustics are reproduced from the measured data using a loudspeaker array and the \gls{nls} method. Section \ref{sec:review2024:nassau_real_time_auralization} addresses the implementation of a real-time auralization system by first reducing the computational complexity of the sound field synthesis using a low-rank approximation of the synthesis filters, followed by the implementation of the acoustic feedback cancelation method discussed in Section \ref{sec:review2024:real_time_auralization}. Finally, Section \ref{sec:review2024:nassau_perceptual_evaluation} presents the perceptual evaluation of the acoustics of the Nassau Chapel using objective acoustic metrics.

\subsection{Room Acoustic Acquisition}
\label{sec:review2024:nassau_room_acoustic_acquisition}

\noindent A series of room acoustic measurements was carried out in the Nassau Chapel to capture the sound field in the space during a historical ceremony or service when the choir is presumed to have been positioned on the choir loft, singing towards the chapel floor. The measurements took place during the night of February 27$^{\textrm{th}}$ 2020, during a time when the chapel was undergoing renovations for the installation of a new exhibition. During the measurements, the chapel was empty and in an unfurnished state. 

\begin{figure}[t]
    \centering
    \includegraphics[width=\columnwidth]{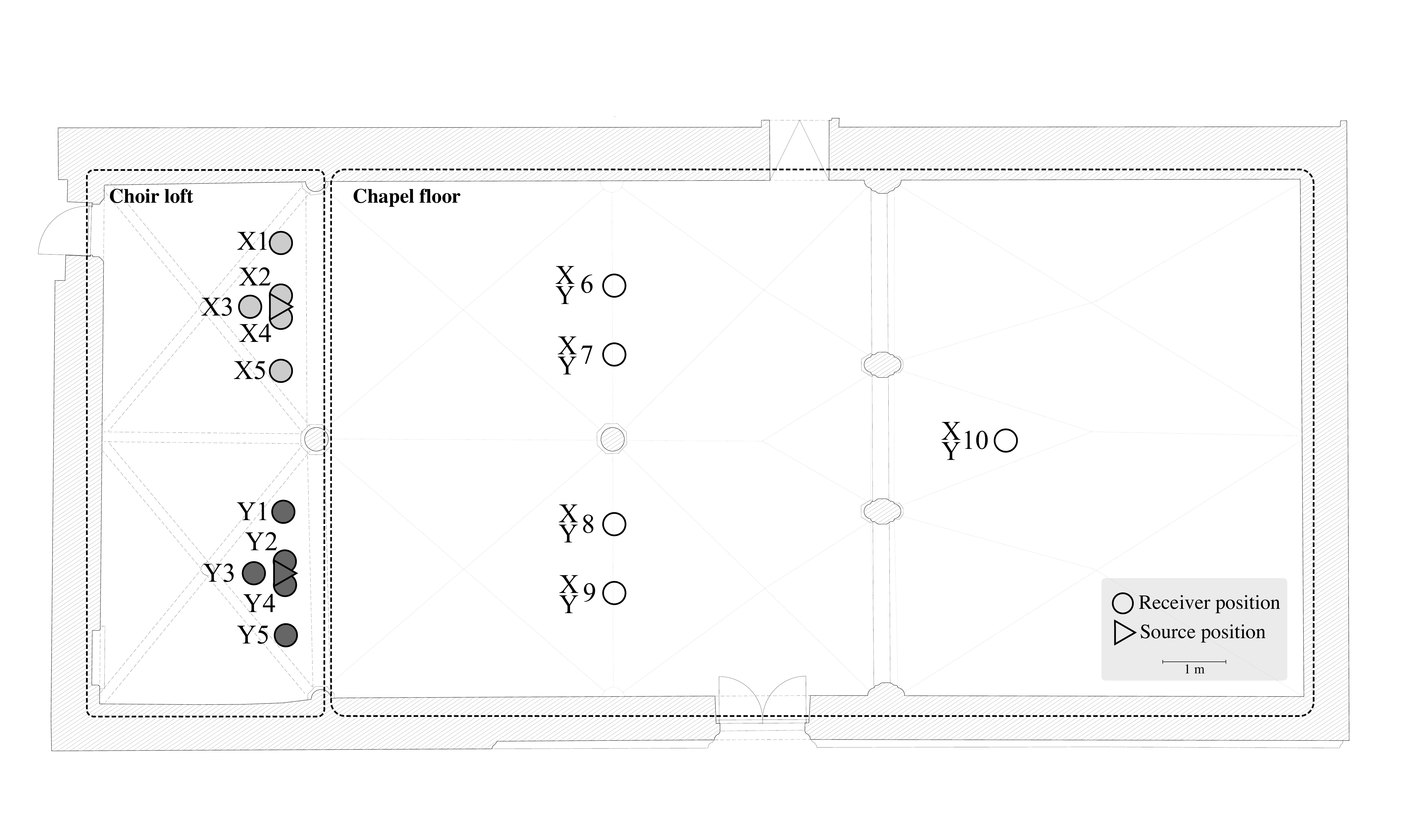}
    \caption{Floor plan of the Nassau Chapel containing the microphone array positions, denoted as circles, and the loudspeaker positions, with the direction of the loudspeaker indicated by the triangle. Two sets of measurements were taken: one with the loudspeaker on the left side of the choir loft (light gray), and the other with the loudspeaker on the right side of the choir loft (dark gray). The microphone array positions were positioned around each loudspeaker position in the choir loft. The microphone positions in the chapel floor are common for both sets of measurements. The measurements are labeled X$1$ to X$10$ for measurements made using the left-side loudspeaker and Y$1$ to Y$10$ for the right-side loudspeaker.}
    \label{fig:review2024:nassau_chapel_floorplan}
\end{figure}

The acoustic measurement setup consisted of a \textit{Genelec 8030C} loudspeaker and a \textit{GRAS 50VI-1 Vector Intensity Probe}, a microphone array consisting of six omnidirectional microphones arranged in a Cartesian grid with a spacing of $25$ mm. The sound levels of each microphone were calibrated before the measurements using a \textit{Bruel \& Kjaer 4230 Sound Level Calibrator}. A floor plan of the Nassau Chapel containing the loudspeaker and microphone array positions is shown in Fig. \ref{fig:review2024:nassau_chapel_floorplan}. Two sets of measurements were taken, one with the loudspeaker on the left side of the choir loft, and the other with the loudspeaker on the right side of the choir loft. For each loudspeaker position in the choir loft, five microphone array positions were chosen around the loudspeaker position to capture the sound field that the choir would have experienced. The acoustic sound field was also captured at the chapel floor during both sets of measurements at five positions. A photograph of the Nassau Chapel with the loudspeaker placed on the right side of the choir loft is shown in Fig. \ref{fig:review2024:nassau_chapel_view}.

The room acoustic measurements were carried out by playing back an \gls{ess} signal \cite{farina2000simultaneous} from the loudspeaker and recording the acoustic response of the chapel with the microphone array. The measurement sampling rate was set at $192$ kHz, and the \gls{ess} signal had a duration of $15$ seconds and a frequency range up to $96$ kHz. The sweeps were repeated five times with an idle time of ten seconds between each sweep to allow the acoustic response of the chapel to decay to a sufficiently low level. After the measurements took place, the recorded responses were deconvolved to obtain the \gls{rir} at each microphone and loudspeaker position.

\subsection{Sound Field Analysis}
\label{sec:review2024:nassau_sound_field_analysis}

\begin{figure}[t]
    \centering
    \includegraphics[width=\columnwidth]{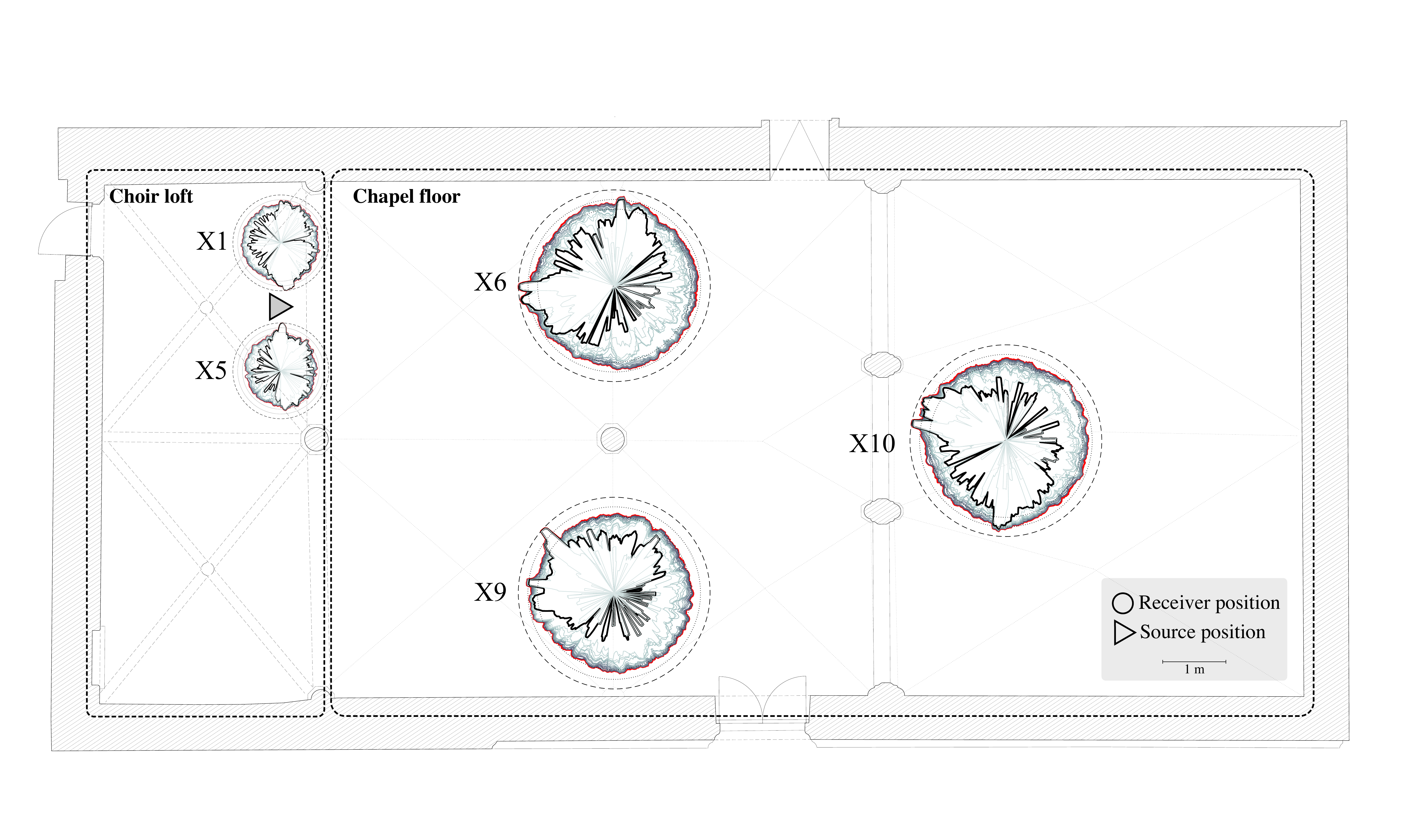}
    \caption{Visualization of the lateral spatio-temporal response at different receiver positions in the Nassau Chapel, with the loudspeaker on the left side of the choir loft. Each line in the spatio-temporal response represents a time instant from $10$ ms to $200$ ms, with a time-increment of $10$ ms. The bold black line represents the cumulative energy at $30$ ms, while the outer red line represents the cumulative energy at $200$ ms. The dashed circle represents the maximum energy of the spatio-temporal response. Each dotted circle represents $-6$ dB of the maximum energy. The energy curves are smoothed with a five-degree angular sliding window average.}
    \label{fig:review2024:nassau_spatial_analysis}
\end{figure}

\noindent The \gls{rir} data obtained from the measurements in the Nassau Chapel were analyzed using the \gls{sdm} to obtain a spatio-temporal representation of the sound field in the chapel. The analysis was carried out for the measurements taken with the loudspeaker on the left side of the choir loft. The spatio-temporal response at different receiver positions in the chapel is shown in Fig. \ref{fig:review2024:nassau_spatial_analysis}. The spatio-temporal response shows the cumulative energy of the sound field at different time instants from $10$ ms to $200$ ms, with a time increment of $10$ ms. The bold black line represents the cumulative energy at $30$ ms, while the outer red line represents the cumulative energy at $200$ ms. The dashed circle represents the maximum energy of the spatio-temporal response, while each dotted circle represents $-6$ dB of the maximum energy. The energy curves are smoothed with a five-degree angular sliding window average. The spatio-temporal response shows the lateral energy distribution of the sound field in the chapel at different time instants, providing insights into the spatial characteristics of the sound field.

It can be seen from the spatio-temporal responses that the cumulative spatial energy distribution at $30$ ms is more diffuse at the choir loft compared to the chapel floor. This is due to the superposition of a large number of early reflections at the choir loft due to the proximity of the walls and ceiling. The energy distribution at the chapel floor is sparser, with the energy arriving directly from the loudspeaker, and the early reflections taking longer to arrive due to the increased distance from the walls and ceiling. From this, we can infer that the choir experiences more noticeable early reflections at the choir loft compared to the chapel floor, which can affect the perception of the sound field in the chapel. The perceptual impact of this spatial energy distribution can be further analyzed through perceptual evaluation, which is discussed in Section \ref{sec:review2024:nassau_perceptual_evaluation}.

\subsection{Sound Field Synthesis}
\label{sec:review2024:nassau_sound_field_synthesis}

\begin{figure}
    \centering
    \includegraphics[width=\columnwidth]{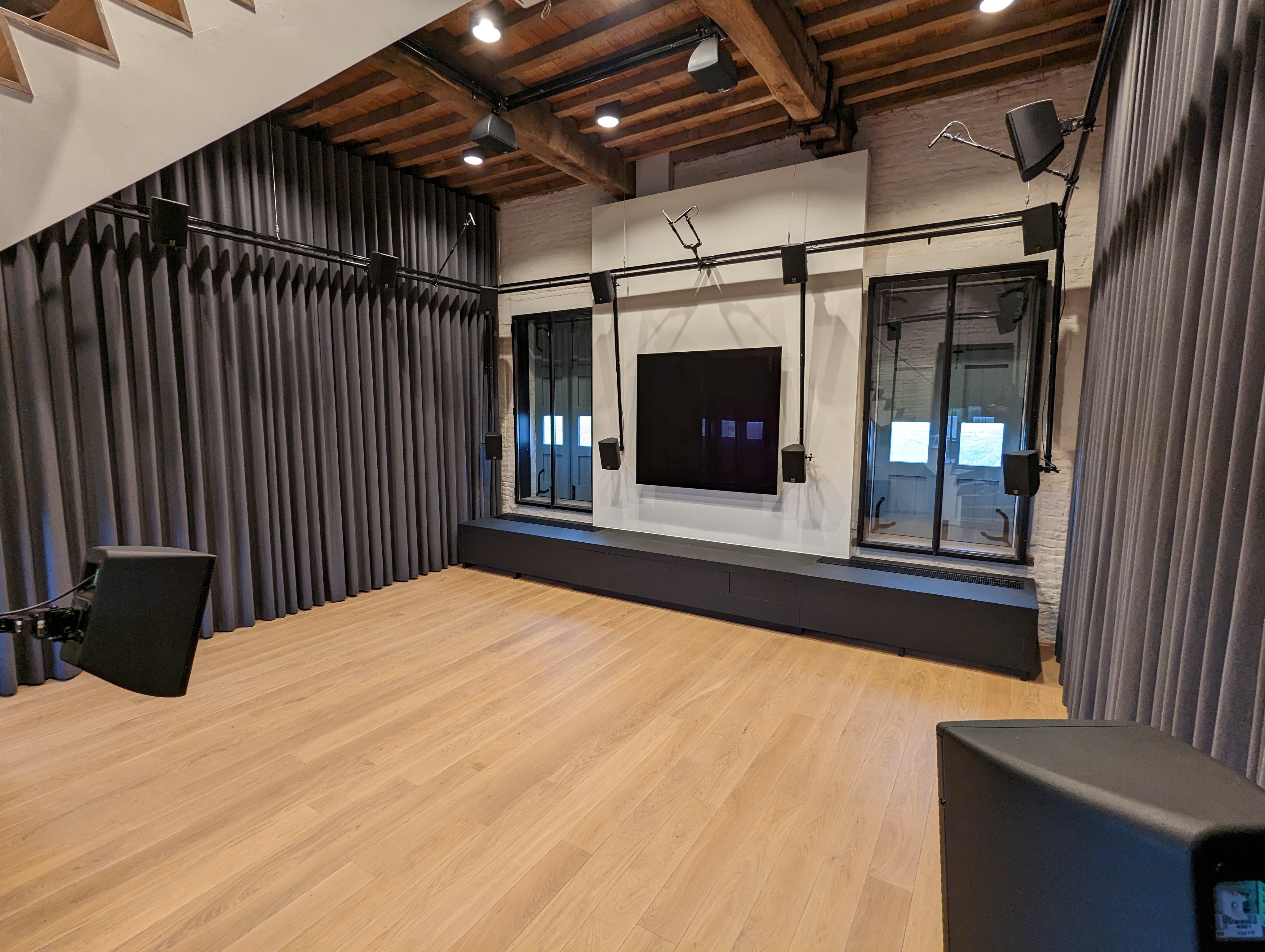}
    \caption{The Alamire Interactive Lab at the Library of Voices in Leuven, Belgium. The lab is equipped with a loudspeaker array containing 24 loudspeakers (Martin Audio CDD6) and six microphones (Audio Technica AT4051b) surrounding the loudspeaker array. The loudspeakers and microphones interface with the computer system through a Dante audio-over-Ethernet protocol to enable low-latency audio streaming between the computer system and the loudspeaker array.}
    \label{fig:review2024:nassau_alamirelab}
\end{figure}

\noindent The measurements and spatial analysis of the Nassau Chapel's acoustics were used to reproduce the acoustics of the chapel in the \gls{ail} at the Library of Voices in Leuven, Belgium. The \gls{ail} is a rehearsal space that is used to study and perform early music, and is equipped with a loudspeaker array containing 24 loudspeakers (Martin Audio CDD6) and six microphones (Audio Technica AT4051b) surrounding the loudspeaker array. The loudspeakers and microphones interface with the computer system through a Dante audio-over-Ethernet protocol to enable low-latency audio streaming between the computer system and the loudspeaker array. The interactive rehearsal space at the \gls{ail} is shown in Fig. \ref{fig:review2024:nassau_alamirelab}.

The reproduction of the Nassau Chapel's acoustics was carried out using \gls{nls}, which maps each image-source from the \gls{sdm} analysis to the closest loudspeaker in the loudspeaker array based on the \gls{doa} of each image-source. This results in a set of synthesis filters that can be convolved with an input signal and played back through the loudspeaker array to reproduce the sound field of the Nassau Chapel. The perceptual evaluation of the reproduced sound field is discussed in Section \ref{sec:review2024:nassau_perceptual_evaluation}.

\subsection{Real-time Auralization}
\label{sec:review2024:nassau_real_time_auralization}

\noindent The Nassau Chapel currently serves as an exhibition space in the Royal Library of Belgium and is therefore not accessible for rehearsals or performances. To enable early music performers to interact with the acoustics of the chapel during rehearsals, a real-time auralization system was implemented in the \gls{ail}.

Real-time auralization in multichannel playback systems requires a high computational cost to convolve the input signal captured by the microphones with all the synthesis filters, one filter for each loudspeaker in the reproduction setup. Since the Nassau Chapel is a highly reverberant space, the synthesis filters contain many filter taps, resulting in a high computational cost that requires a powerful processor to handle the convolution operations in real-time. When the processor is not able to handle the computational load in real-time, the system introduces latency which can degrade the quality of the auralization and make it unsuitable for real-time applications. To lower the computational cost of convolution with the synthesis filters, a low-rank approximation of the filters was applied to reduce the number of filter taps and therefore the computational complexity of the convolution operation. In \cite{rosseel2024low}, the authors applied a low-rank approximation to the synthesis filters, achieving a significant reduction in computational complexity while maintaining key perceptual features of the sound field.

Additionally, real-time auralization using loudspeakers and microphones to reproduce the sound field is prone to acoustic feedback, which occurs when the output signal from the loudspeaker array is picked up by the microphones and fed back into the auralization system. Acoustic feedback causes instability in the system and degrades the quality of the auralization by introducing artifacts such as howling or unwanted echo. To mitigate the effects of acoustic feedback in the \gls{ail}, the acoustic feedback cancelation method presented in \cite{abel2018feedback} was implemented for the real-time auralization system.

\subsection{Perceptual Evaluation}
\label{sec:review2024:nassau_perceptual_evaluation}

The room acoustic measurements were analyzed to extract various room acoustic parameters, including the Reverberation Time ($T_{20}$), \gls{edt}, Clarity ($C_{80}$), Definition ($D_{50}$), Center time ($T_s$), and \gls{drr}. The parameters were calculated in accordance with the ISO:3382-1 standard \cite{ISO3382-1}. The objective room acoustic parameters were calculated for each microphone position in the chapel and are presented in Table \ref{tab:review2024:objective_params}. The results show that the reverberation time in the Nassau Chapel is relatively long, with values ranging from $5.32$ to $6.20$ seconds. Although unusual for a chapel of this size, the reverberation time is expected to be higher since the measurements were conducted in an empty chapel due to ongoing renovations, rather than when the chapel is furnished. Moreover, a clear distinction can be observed between the measurements made on the choir loft (1-5), and the measurements made on the chapel floor (6-10). The measurements on the choir loft have a lower reverberation time and center time compared to the measurements on the chapel floor. This is expected since the choir loft is positioned closer to the chapel's ceiling. Moreover, the clarity, definition, energy and DDR are higher on the choir loft compared to the chapel floor, confirming that the sound field is sparser and less reverberant on the choir loft.

The reproduced acoustics of the Nassau Chapel in the \gls{ail} were evaluated using the same objective room acoustic parameters to compare the psychoacoustic characteristics of the reproduced sound field with the measured data. The objective parameters were calculated for the auralization of position X7 in the Nassau Chapel and are presented in Table \ref{tab:review2024:objective_params}. The results show that the objective parameters of the auralization of position X7 are similar to the measured data, with most parameters of the auralization falling within 1 \gls{jnd} of the measured data. This indicates that the auralization system is able to reproduce the acoustics of the Nassau Chapel with the same psychoacoustic characteristics as the measured data. The auralization does show a significantly lower \gls{drr} compared to the measured data, indicating that the direct-path component of the sound field is less pronounced in the auralization compared to the measured data. Further investigation is needed to determine the cause of this discrepancy and to improve the auralization system to better match the measured data.

\begin{table*}[ht]
    \centering
    \caption{Objective acoustic parameters for all measurement positions in the Nassau Chapel, averaged from five consecutive measurements at each position. Objective parameters were also calculated for the Alamire Interactive Lab and the auralization of position X7 in the Nassau Chapel. The objective parameters were calculated according to single number frequency averaging in the 500 Hz to 1 kHz octave band, in accordance with the ISO:3382-1 standard \cite{ISO3382-1}.}
\begin{tabular}{cccccccc}
        \toprule
        Position                & $T_{20}$ [s] & EDT [s] & $C_{80}$ [dB] & $D_{50}$ [dB] & $T_s$ [s] & DRR [dB] \\
        \midrule
        X1                      & 5.88         & 5.22    & -0.18         & 0.43          & 0.26      & -6.62    \\
        Y1                      & 5.79         & 5.47    & 0.32          & 0.46          & 0.25      & -6.30    \\\\[-0.8em]
        X2                      & 5.40         & 4.27    & 5.56          & 0.75          & 0.11      & -0.14    \\
        Y2                      & 5.32         & 3.76    & 6.02          & 0.77          & 0.10      & 0.32     \\\\[-0.8em]
        X3                      & 5.83         & 5.25    & 0.15          & 0.44          & 0.26      & -5.12    \\
        Y3                      & 5.84         & 5.40    & 0.72          & 0.47          & 0.24      & -5.36    \\\\[-0.8em]
        X4                      & 5.49         & 4.26    & 6.20          & 0.79          & 0.10      & 0.63     \\
        Y4                      & 5.42         & 3.96    & 6.22          & 0.79          & 0.10      & 0.01     \\\\[-0.8em]
        X5                      & 5.87         & 5.84    & 0.12          & 0.44          & 0.27      & -6.43    \\
        Y5                      & 5.84         & 5.41    & 0.08          & 0.45          & 0.25      & -6.47    \\\\[-0.8em]
        X6                      & 6.07         & 5.85    & -6.74         & 0.12          & 0.42      & -11.53   \\
        Y6                      & 6.06         & 6.26    & -6.13         & 0.14          & 0.43      & -15.41   \\\\[-0.8em]
        X7                      & 6.15         & 6.11    & -6.64         & 0.12          & 0.44      & -11.54   \\
        Y7                      & 6.04         & 6.02    & -5.75         & 0.13          & 0.43      & -13.96   \\\\[-0.8em]
        X8                      & 6.18         & 6.23    & -7.86         & 0.09          & 0.47      & -15.35   \\
        Y8                      & 6.07         & 6.06    & -7.43         & 0.10          & 0.45      & -15.38   \\\\[-0.8em]
        X9                      & 6.17         & 6.14    & -6.11         & 0.15          & 0.43      & -14.84   \\
        Y9                      & 6.05         & 5.77    & -6.15         & 0.14          & 0.42      & -11.68   \\\\[-0.8em]
        X10                     & 6.05         & 6.11    & -5.84         & 0.14          & 0.41      & -15.64   \\
        Y10                     & 6.20         & 6.21    & -6.82         & 0.11          & 0.44      & -11.31   \\\\[-0.8em]
        Alamire Interactive Lab & 0.54         & 0.55    & 9.72          & 0.76          & 0.03      & -7.01    \\
        Auralization of X7      & 5.98         & 6.42    & -6.87         & 0.12          & 0.48      & -24.32   \\
        \bottomrule
    \end{tabular}
    \label{tab:objective_params}
\end{table*}
  \section{Conclusions}
\label{sec:review2024:conclusion}
\glsresetall

In this paper, we have presented a comprehensive review regarding the state of research in the field of acoustic preservation of historical \gls{hws} through auralization. Acoustic preservation through auralization is a multidisciplinary field that integrates acoustics, signal processing, and psychoacoustics to faithfully recreate the acoustic characteristics of these spaces. Over recent decades, significant advancements have been made, particularly in the development of methods for the acquisition, analysis, synthesis, and evaluation of historical worship space acoustics. The literature reviewed has focused on various stages of the auralization process, including room acoustic measurement, sound field analysis, synthesis techniques, real-time auralization, and perceptual assessment. Furthermore, applications of these methods to historical \gls{hws} have been explored in depth.

While established methods such as \gls{hoa}, \gls{shd}, and \gls{wfs} have proven effective in concert hall acoustics, their application to historical \gls{hws} has been limited. Future research should focus on applying and adapting these techniques to the analysis and auralization of historical \gls{hws}. Additionally, the development of interactive auralization tools that allow musicians, researchers, and other stakeholders to experience and interact with the acoustics of historical \gls{hws} in real-time is an important avenue for future work. Overcoming challenges related to computational complexity, latency, and acoustic feedback will be crucial for achieving seamless, interactive simulations.

Additionally, the evaluation of the perceptual impact of historical worship space acoustics and auralization systems has primarily relied on objective acoustic parameters to evaluate the psychoacoustic effects on listeners. However, to gain a better understanding of the subjective experience of listeners and how the auralized spaces are perceived, future research should incorporate subjective and sensory evaluation methods.

Several reviewed methods were illustrated in a case study of the Nassau Chapel, where auralization techniques were applied to preserve and valorize the chapel's acoustics through real-time interactive auralization. This case study demonstrated the potential of interactive auralization tools to support preservation efforts, offering musicians the opportunity to rehearse and perform in a virtual recreation of the chapel's original acoustics.

This review paper serves as a reference for researchers, acousticians, architects, and cultural heritage professionals, providing a framework for future preservation projects. An outline of the general process for acquiring, analyzing, synthesizing, and evaluating the acoustics of these spaces has been presented, which can help guide future research in the field.

In conclusion, the preservation and auralization of historical \gls{hws} is an important interdisciplinary field of research that combines acoustics, signal processing, and psychoacoustics to safeguard and recreate the original acoustic characteristics of these spaces. The reviewed research offers a solid foundation for further exploration, pointing to key areas for future development. 
\section*{Acknowledgements}
\noindent The authors express their gratitude to the Alamire Foundation and Ann Kelders for their invaluable information and late night access to the Nassau Chapel. Additionally, special thanks are extended to Randall Ali for assisting with the acoustic measurements conducted in the Nassau Chapel. This research work was carried out at the ESAT Laboratory of KU Leuven, in the frame of KU Leuven internal funds C3/23/056 “HELIXON: Hybrid, efficient, and liquid interpolation of sound in extended reality” and C14/21/075 "A holistic approach to the design of integrated and distributed digital signal processing algorithms for audio and speech communication devices”, FWO SBO Project The sound of music - Innovative research and valorization of plainchant through digital technology" (S005319N), FWO Large-scale research infrastructure "The Library of Voices - Unlocking the Alamire Foundation's Music Heritage Resources Collection through Visual and Sound Technology" (I013218N), and SBO Project "New Perspectives on Medieval and Renaissance Courtly Song" (S005525N). The research leading to these results has received funding from the Flemish Government under the AI Research Program and from the European Research Council under the European Union's Horizon 2020 research and innovation program / ERC Consolidator Grant: SONORA (no. 773268). This paper reflects only the authors' views and the Union is not liable for any use that may be made of the contained information.

\printglossary[type=\acronymtype, nonumberlist]

\end{document}